# From Stellar Death to Cosmic Revelations: Zooming in on Compact Objects, Relativistic Outflows and Supernova Remnants with AXIS

Safi-Harb, S.[1], Burdge K. B.[2], Bodaghee, A.[3], An, H.[4], Guest, B.[5,6,7], Hare, J.[6,7,8], Hebbar, P.[9], Ho, W. C. G.[10], Kargaltsev, O.[11], Kirmizibayrak[12], D., Klingler, N.[13,6,7], Nynka, M.[14], Reynolds, M. T.[15,16], Sasaki, M.[17], Sridhar, N.[18,19], Vasilopoulos, G.[20,21], Woods, T. E.[1], Yang, H.[11], Heinke, C.[9], Kong, A.[22], Li, J.[23], MacMaster, A.[1], Mallick, L.[1,24,19,*], Treyturik, C.[1], Tsuji, N.[25], Binder, B.[26], Braun, C.[1], Chang, H.-K.[27], Chatterjee, A.[1], Ferrand, G.[1,28], Holland-Ashford, T.[29], Ng, C.-Y.[30], Plotkin, R.[31], Romani, R.[32], Zhang, S.[33], for the *AXIS* Compact Objects-Supernova Remnants Science Working Group

1. Department of Physics & Astronomy, University of Manitoba, Winnipeg, MB R3T 2N2, Canada; samar.safi-harb@umanitoba.ca
2. Department of Physics, Massachusetts Institute of Technology, Cambridge, MA 02139, USA; kburdge@mit.edu
3. Department of Chemistry, Physics and Astronomy, Georgia College and State University, 231 W. Hancock St., Milledgeville, GA 31061, USA
4. Department of Astronomy and Space Science, Chungbuk National University, Cheongju, 28644, Republic of Korea
5. Department of Astronomy, University of Maryland, College Park, MD 20742, USA
6. Astrophysics Science Division, NASA Goddard Space Flight Center, 8800 Greenbelt Rd, Greenbelt, MD 20771, USA
7. Center for Research and Exploration in Space Science and Technology, NASA/GSFC, Greenbelt, MD 20771, USA
8. The Catholic University of America, 620 Michigan Ave., N.E. Washington, DC 20064, USA
9. Physics Dept., CCIS 4-183, University of Alberta, Edmonton, AB, T6G 2E1, Canada
10. Department of Physics and Astronomy, Haverford College, 370 Lancaster Avenue, Haverford, PA, 19041, USA
11. Department of Physics, The George Washington University, 725 21st Street NW, Washington, DC 20052, USA
12. Department of Physics and Astronomy, University of British Columbia; Vancouver, BC, V6T 1Z1, Canada
13. Center for Space Sciences and Technology, University of Maryland, Baltimore County, Baltimore, MD, 21250, USA
14. Kavli Institute For Astrophysics and Space Research, Massachusetts Institute of Technology, Cambridge, MA, USA
15. Department of Astronomy, The Ohio State University, 140 W. 18th Ave., Columbus, OH 43210, USA
16. Department of Astronomy, University of Michigan, 1085 S. University Ave., Ann Arbor, MI 48109, USA
17. Dr. Karl Remeis Observatory, Erlangen Centre for Astroparticle Physics, Friedrich-Alexander-Universität Erlangen-Nürnberg, Sternwartstraße 7, 96049 Bamberg, Germany
18. Department of Astronomy and Columbia Astrophysics Laboratory, Columbia University, 550 W 120th St, New York, NY 10027, USA
19. Cahill Center for Astronomy and Astrophysics, California Institute of Technology, Pasadena, CA 91125, USA
20. Department of Physics, National and Kapodistrian University of Athens, University Campus Zografos, GR 15783, Athens, Greece
21. Institute of Accelerating Systems & Applications, University Campus Zografos, Athens, Greece
22. Institute of Astronomy, National Tsing Hua University, Hsinchu 30013, Taiwan
23. Purple Mountain Observatory, Chinese Academy of Sciences, 10 Yuanhua Road, Nanjing 210023, People's Republic of China
24. Canadian Institute for Theoretical Astrophysics, University of Toronto, 60 St George Street, Toronto, Ontario M5S 3H8, Canada

25   Faculty of Science, Kanagawa University, 3-27-1 Rokukakubashi, Kanagawa-ku, Yokohama-shi, Kanagawa 221-8686, Japan

26   Department of Physics and Astronomy, California State Polytechnic University, Pomona, 3801 W. Temple Ave., Pomona, CA 91768, USA

27   Institute of Astronomy, National Tsing Hua University. Hsinchu 300044, Taiwan

28   RIKEN Interdisciplinary Theoretical and Mathematical Sciences program (iTHEMS), Wako, Saitama 351-0198, Japan

29   Center for Astrophysics, Harvard & Smithsonian, 60 Garden St, Cambridge MA 02138, USA

30   Department of Physics, The University of Hong Kong, Pokfulam Rd., Hong Kong

31   Department of Physics, University of Nevada, Reno, NV, USA

32   KIPAC/Dept. of Physics, Stanford University, Stanford CA, 94305, USA

33   Bard College Physics Program, 30 Campus Road, Annandale-on-Hudson, NY 12504, USA

*  CITA National Fellow

**Abstract:** Compact objects and supernova remnants provide nearby laboratories to probe the fate of stars after they die, and the way they impact, and are impacted by, their surrounding medium. The past five decades have significantly advanced our understanding of these objects, and showed that they are most relevant to our understanding of some of the most mysterious energetic events in the distant Universe, including Fast Radio Bursts and Gravitational Wave sources. However, many questions remain to be answered. These include: What powers the diversity of explosive phenomena across the electromagnetic spectrum? What are the mass and spin distributions of neutron stars and stellar mass black holes? How do interacting compact binaries with white dwarfs - the electromagnetic counterparts to gravitational wave LISA sources - form and behave? Which objects inhabit the faint end of the X-ray luminosity function? How do relativistic winds impact their surroundings? What do neutron star kicks reveal about fundamental physics and supernova explosions? How do supernova remnant shocks impact cosmic magnetism? This plethora of questions will be addressed with *AXIS* - the Advanced X-ray Imaging Satellite - a NASA Probe Mission Concept designed to be the premier high-angular resolution X-ray mission for the next decade. AXIS, thanks to its combined (a) unprecedented imaging resolution over its full field of view, (b) unprecedented sensitivity to faint objects due to its large effective area and low background, and (c) rapid response capability, will provide a giant leap in discovering and identifying populations of compact objects (isolated and binaries), particularly in crowded regions such as globular clusters and the Galactic Center, while addressing science questions and priorities of the US Decadal Survey for Astronomy and Astrophysics (Astro2020).

*This White Paper is part of a series commissioned for the AXIS Probe Concept Mission; additional AXIS White Papers can be found at the AXIS website with a mission overview here.*

## Contents

**List of Figures**

## 1. Introduction

The X-ray sky has provided invaluable insight into the formation and evolution of compact objects, supernova remnants, and pulsar wind nebulae. This incredible success has been driven by many X-ray missions however Chandra in particular has been instrumental in providing the high angular resolution studies needed to characterize dense regions and perform astrometric measurements. In this white paper, we discuss the enormous potential of the proposed Advanced X-ray Imaging Satellite (AXIS) [248,306] to capitalize on this legacy and make the next groundbreaking advances in the study of compact objects and supernova remnants.

With relatively uniformly high angular resolution over its entire field of view, *AXIS* is uniquely poised to conduct high angular resolution surveys over dense regions in the Galactic Plane and Bulge, delivering images capable of resolving almost all X-ray sources with exceptional sensitivity. Futhermore, *AXIS* is uniquely suited to study diffuse objects such as supernova remnants and pulsar wind nebulae, where it can use its powerful angular resolution to resolve the detailed structure of these extended objects over the entire FoV. This capability also allows *AXIS* to make exquisite proper motion measurements of the compact objects within these remnants by using a precise astrometric solution derived from X-ray sources throughout its entire field of view. These are just a few of the many advances we highlight in this white paper.

*AXIS* will address key questions highlighted in the 2020 Decadal Survey of Astronomy & Astrophysics (Astro2020)[1][250] regarding new messengers and new physics with these objects, and provide new windows on the dynamic universe by delivering surveys with high temporal resolution that complement time-domain surveys at other wavelengths. In this paper, we discuss a proposed Galactic plane survey (GPS) with *AXIS* and how it will shed light on the origin of Type Ia supernovae. The GPS will be the largest, most sensitive, high-resolution X-ray map of the Galactic plane ever made, providing a unique view on diverse scientific topics related to both compact objects and diffuse emission. We also highlight a broad list of guest investigator science where *AXIS* will be impactful, including studies of the Galactic Center, open and globular clusters, isolated neutron stars, accreting binaries, pulsar wind nebulae, supernovae remnants, and outflows from binaries.

## 2. Galactic Plane Survey

The inner quadrant of the Milky Way is densely packed with massive stars residing among the wide lanes of obscuring gas and dust that form the spiral arms. When a massive star goes supernova (SN), its core forms a compact object (CO) such as a neutron star (NS) or a black hole (BH). Low-mass stars like the Sun end their lives as a planetary nebula where the outer layers are shed, leaving behind a white dwarf (WD). Thus, the inner galactic plane hosts the greatest concentration of systems with COs, e.g., X-ray binaries (XRBs) with low-mass or high-mass stellar companions (LMXBs and HMXBs, respectively), cataclysmic variables (CVs), ultra-compact WD binaries, rotation-powered pulsars (RPPs), pulsar and magnetar wind nebulae (PWNe and MWNe), and supernova remnants (SNRs). X-ray photons from these systems are messengers of matter subject to extreme gravitational and electromagnetic fields. In addition, some of these systems could form inspiraling binaries which are precursors to the types of gravitational wave events detectable by LIGO/Virgo, and eventually *LISA* [e.g., 5,212,267].

X-ray surveys by *Chandra* detected 9017 sources in a $2° \times 0.8°$ section of the Galactic center [247]. The deepest exposures ($\sim$1 Ms) were centered on Sgr A$^*$ where the source density approached 10 sources per arcmin$^2$ with a minimum luminosity of $L \geq 3 \times 10^{31}$ erg s$^{-1}$ in the 0.5–8 keV energy range. A similar

---

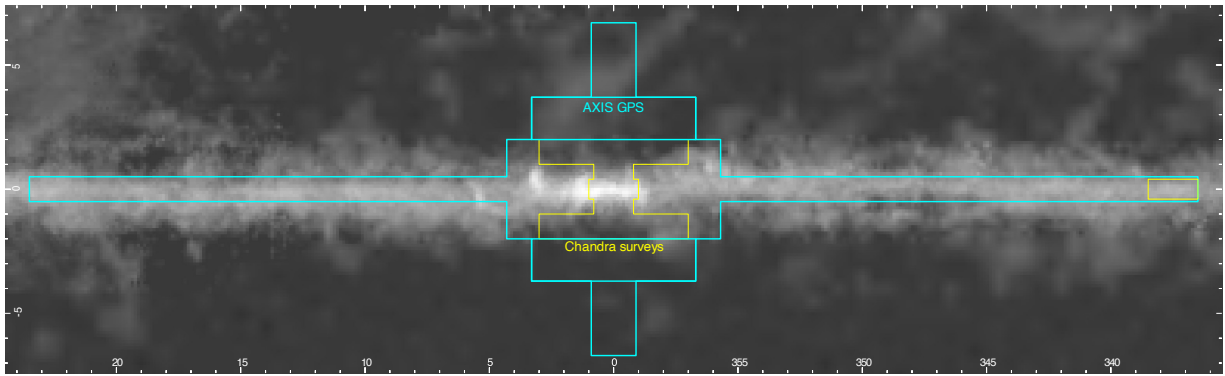

**Figure 1.** CO map of Dame et al. [76] overlaid with the outlines of the *AXIS* GPS footprint in cyan and the existing *Chandra* surveys in yellow. The *AXIS* GPS covers an area of 104 deg$^2$ with a sensitivity limit $\sim 10^{-15}$ erg cm$^{-2}$ s$^{-1}$ for a $5\sigma$ detection ensuring the discovery of a million new X-ray sources. Thanks to its low background and narrow PSF, *AXIS* will locate these sources with arc-second precision which is required for multi-wavelength counterpart-matching in these crowded regions.

survey of the Norma Arm uncovered 1415 sources [99], while a shallower, but wider (2 × 6° × 1°), survey of the galactic bulge found 1640 sources [175]. The central parsecs around Sgr A* includes sources with non-thermal spectra distinct from the thermal spectra of active binaries (ABs), dwarf novae (DNe), and non-magnetic CVs that are the dominant X-ray populations in the center, bulge, and plane. Among these, at least 6 NS-LMXB systems and 18 BH-LMXB candidates have been proposed [e.g., 204,240]. Their spatial distribution and luminosity function suggest a population of hundreds of quiescent BH- and NS-LMXBs, with decades-long periods of inactivity lurking below the *Chandra* sensitivity limit [109,137].

In addition to these point sources, the inner galaxy contains large-scale diffuse background emission in the X-rays and gamma-rays: e.g., the hot gas surrounding Sgr A* [e.g., 245]; the Galactic Chimneys [288]; the *Fermi* Bubbles [e.g., 348] and GeV excess [e.g., 8]; the 511-keV line from positron annihilation [e.g., 335,336]; and the Galactic Ridge X-ray Emission [GRXE: e.g., 305]. In some cases, the diffuse emission is the result of the combined emission from different populations of resolved and unresolved point sources, each with its own luminosity range and continuum temperature. For example, ∼80% of the GRXE around 6–8 keV can be attributed to thousands of accreting WDs such as CVs and DNe [e.g., 79,274,304,380,396]. Faint, coronally-active stars and ABs are believed to produce the ∼1 keV low-temperature emission [e.g., 404], while the more luminous population of magnetic CVs [158] and intermediate-polar CVs [138] may explain the emission seen around 15 keV [359,404]. Rotation-powered pulsars (RPPs) are predicted to be present in significant numbers in the Galactic center and could be responsible for the GeV excess [8], and possibly, in combination with flaring stars and XRBs, the origin of the 511-keV signal as well [336]. However, only one such object has been confirmed within 20 pc of Sgr A*: the pulsar wind nebula PSR J0002+6216 ("Cannonball"). The lack of RPPs in the Galactic center is puzzling; while simulations show that pulsars can be rapidly ejected from their birthplace due to natal kicks [41], the existence of pulsars within globular clusters suggests that there should be pulsars in the Galactic center as well. Additionally, the Galactic center environment encourages the production of magnetars, instead of normal-B neutron stars, and so the NS population could include a large fraction of magnetars [84]. The detection of the missing magnetar and pulsar population in the central regions of the galaxy, in addition to the first discovered magnetar (SGR J1745–2900, ∼2.4$''$ away from Sgr A*) and the Cannonball pulsar, would shed light on this hidden population's abundance (i.e., how massive stars evolve) and distribution (i.e., the energetics of SNe).

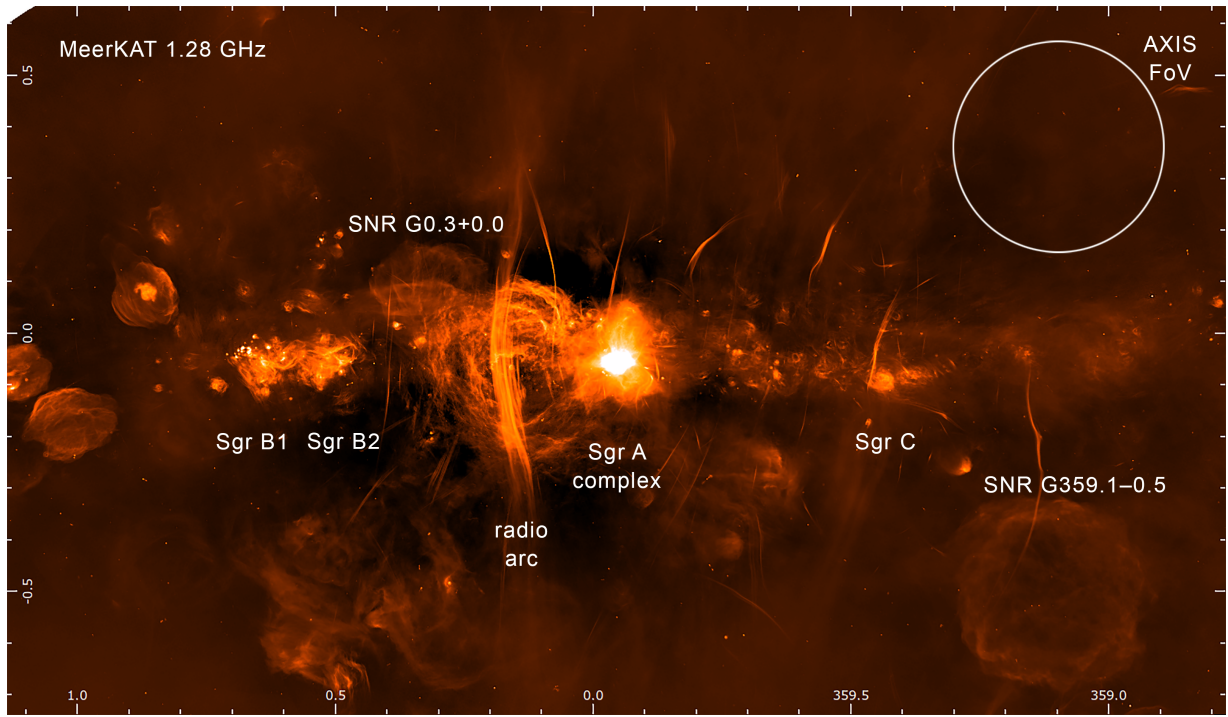

**Figure 2.** MeerKAT image in 1.28 GHz showing the complex structure of the diffuse radio emission in the GC [152]. The size of the *AXIS* field-of-view is shown as a white circle in the upper right. This field will be fully mapped by the *AXIS* GPS.

Therefore, the primary science question is: **which X-ray populations inhabit the faint end of the X-ray luminosity function and how do they contribute to the diffuse X-ray and gamma-ray background?** Unfortunately, a census of the different X-ray populations that occupy the inner galaxy is incomplete. Existing *Chandra* and *XMM-Newton* surveys form a relatively small patchwork of images with exposure times that are not uniform. Most of the observed area has only limited sensitivity making these surveys biased towards bright X-ray populations. High latitudes of the bulge, where the absorption by dust is diminished, and almost all of the inner disk remains unmapped. The *Chandra* field-of-view (FoV) is narrow which makes mapping a large area impractical. *XMM-Newton* has a wider FoV but its point-spread-function (PSF) is large which hinders the resolving of individual point sources in dense fields. In both of these instruments, the PSF broadens for sources located off axis. Relating the X-ray source to a counterpart at lower energies is crucial for determining which class of system is involved. While telescopes operating in the radio, infrared, and optical bands feature astrometric precision typically <0.1 arcsec, X-ray sources are usually discovered with position uncertainties of $0.6 - 1$ arcsec at the aimpoint of narrow-FoV focusing telescopes or a few arcmin for wide-FoV coded-mask instruments. Given the stellar density in the plane, and especially in the center, such wide X-ray error circles encompass several dozen candidates, any one of which could be the low-energy counterpart.

The *AXIS* Galactic Plane Survey (GPS: Fig. 1) will cover a 104-deg$^2$ swath of the sky centered on the GC extending to longitudes of $\pm 23.5$ deg, i.e., from the Scutum/Sagittarius Arm to the Inner/Norma Arm, with latitudes of $\pm 0.5$ deg. Around 65 deg$^2$ is dedicated to mapping the Bulge up to low-extinction latitudes of $\pm 7$ deg. Its wide FoV and narrow PSF make *AXIS* perfectly suited for creating a wide-field census of the Milky Way's X-ray populations. These observations represent a legacy-level dataset that will complement existing surveys in other wavelengths, as well as overlap with planned surveys from the Rubin and Roman telescopes, and it will serve in the study of all X-ray source classes.

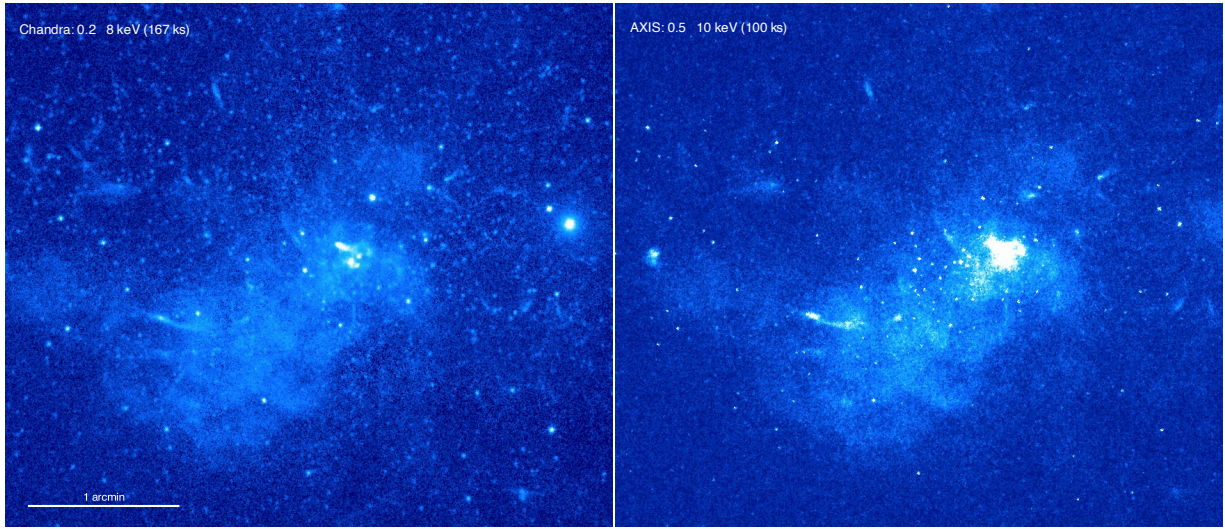

**Figure 3.** The left panel shows the *Chandra* image (ObsID 3392; 167 ks) of the vicinity of Sgr A* and the right panel presents a simulated *AXIS* image of the same region (100 ks). Both images are given in Galactic coordinates with logarithmic scaling. One arcmin is equivalent to 2.3 pc at the Galactic center distance of 8 kpc.

Covering the GPS field outlined in Fig. 1 will require ~1000 tiles assuming the hexagonal dithering of a 24-arcmin-wide circular FoV with 15% overlap between adjacent tiles to prevent gaps. A minimum of 6 ks of exposure is allocated to each tile so that the limiting flux (observed in 0.5–10 keV) for a source detected at significance of $5\sigma$ is $\sim 10^{-15}$ erg cm$^{-2}$ s$^{-1}$ assuming a power-law spectrum with $2 \times 10^{22}$ cm$^{-2}$ and $\Gamma$ = 1.5 (i.e., the same parameters as in [175]). This corresponds to a luminosity $L_X \sim 10^{31}$ erg s$^{-1}$ for a distance of 8 kpc. In the GC, where the exposure times from GO observations are expected to surpass 250 ks, the flux limits will be $10^{-16}$ erg cm$^{-2}$ s$^{-1}$ ($L_X \sim 10^{30}$ erg s$^{-1}$). This means **the GPS will be the most sensitive, high-resolution X-ray map of the Galactic plane ever made with a flux limit 1–2 orders of magnitude below those of *Chandra* and *XMM-Newton*.**

The most interesting regions of the GPS such as the Galactic center, bulge, and tangents to the inner spiral arms, will benefit from deeper exposures thanks to the Guest Observer (GO) program. Figure 3 presents a simulated 100-ks *AXIS* observation of the Galactic center compared with a 167-ks *Chandra* observation of the same field. Figure 4 presents simulated images and flux distributions for sources in a typical region of the bulge. Exposure times of 5 ks, 50 ks, and 250 ks, respectively, will enable the detection of 9%, 45%, and 85% of the faint, unresolved X-ray population. For the plane, bulge, and center, respectively, extrapolation of the *Chandra* log$N$-log$S$ curve [247] with slopes of $-1.0$, $-1.3$, and $-1.5$ to these lower flux limits yields source densities of $3.6 \times 10^4$, $1.5 \times 10^5$, and $3 \times 10^6$ sources deg$^{-2}$.

Within the GPS footprint, over 1 million sources are expected to be discovered along with the re-detection of 35,000–50,000 sources listed in the serendipitous survey catalogs of *Chandra* and *XMM-Newton*. In the center, all known CVs and XRBs with a distance less than 15 kpc will be detected, as well as previously undetected populations of faint CVs and quiescent LMXBs out to 8 kpc. In the bulge and plane, the known population of persistent and active-transient XRBs will be detected through the galaxy; transient systems in quiescence will be observable to a distance of 4–8 kpc. A detection significance of $5\sigma$ ensures that even if most sources emit only a handful of counts, they will have an X-ray position with an error radius of ≲1 arcsec which will restrict the number of candidate counterparts for follow-up studies. Optical/infrared spectroscopy of these candidates, along with basic X-ray timing and spectral information from the GPS observation or from deeper follow-up observations with other telescopes, will help categorize the object into one of the source classes.

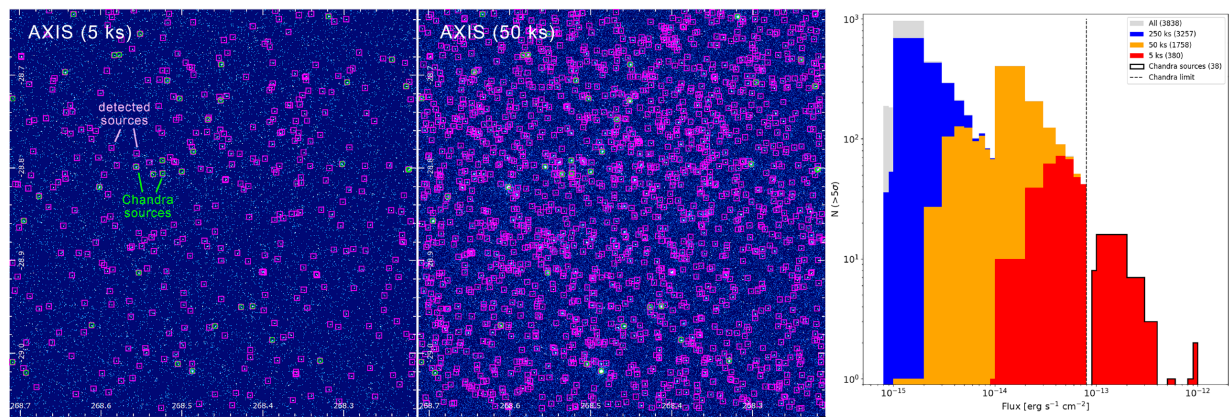

**Figure 4.** Simulated *AXIS* image in 0.5–10 keV for a representative section of the Galactic bulge centered at $(l; b) = (+1.0; -1.5)$ with 5 ks and 50 ks of exposure time (two left-most panels). In addition to the 38 *Chandra* sources (circled in green) within the *AXIS* FoV whose fluxes are given [175], the simulation included 3,800 faint, uniformly-distributed mock sources. All sources were assigned an absorbed power-law spectrum ($N_H = 2 \times 10^{22}$ cm$^{-2}$; $\Gamma = 1.5$) with mock sources given fluxes ranging from the *Chandra* sensitivity limit of $8 \times 10^{-14}$ erg cm$^{-2}$ s$^{-1}$ [175] down to $8 \times 10^{-16}$ erg cm$^{-2}$ s$^{-1}$ according to a log$N$–log$S$ curve with $\alpha = 1.3$ [247]. Magenta squares denote sources detected above $5\sigma$ significance. The right panel shows the flux distribution of detected sources (with numbers in parentheses) for typical exposure times of the GPS ($\sim 5$ ks) and GO programs (50–250 ks). The *Chandra* flux distribution and sensitivity limit are indicated by the thick and dashed lines, respectively.

The data will be valuable for studies of diffuse emission beyond the link with point sources: the regions covered include where the Galactic Chimneys meet the *Fermi* Bubbles, the peak and possible asymmetric profile of the positron annihilation signal, and the molecular clouds and filamentary structures highlighted in recent radio maps by MeerKAT [152]. The tight PSF and low detector background of *AXIS* are ideal for X-ray imaging and spectral analysis of faint, diffuse structures allowing a broad-band study of these features from radio to X-rays (Fig. 2).

*Which CO populations inhabit the faint end of the X-ray luminosity function and how do they contribute to diffuse emission?* The *AXIS* GPS will reveal a crowd of X-ray sources hidden in the background of previous surveys of the Milky Way, which will clarify the origin of different types of diffuse emission: e.g., point-vs.-diffuse emission in the vicinity of Sgr A*; pulsars and the 511-keV line; ABs/CVs and the GRXE; and the *Fermi* Bubbles. All source classes are expected in the GPS, including extragalactic sources, SNRs, and the precursors to GW events, so the GPS data will hold legacy value by complementing and expanding on past (*Spitzer*), present (*Chandra*, *eRosita*), and future (Roman, Rubin) multi-wavelength surveys.

*What can AXIS measure?* The GPS represents a sky area of 104 deg$^2$, inside of which *AXIS* will detect around 1 million new objects with (sub-)arcsec error radii. In the crowded fields of the Galactic center and plane, an error radius smaller than an arcsec is required to identify the likely optical/IR counterparts. Along with the multi-wavelength counterparts, *AXIS* will also provide spectral and timing information helping to narrow down the class to which each source belongs. Thanks to its low detector background, *AXIS* will map the low-surface brightness diffuse emission with a better sensitivity and spatial resolution than previously possible, enabling the location (hence, the energetics and dynamics) of sources responsible for accelerating particles that lead to diffuse emission like the GRXE or positron annihilation signatures, as well as outflows such as the Galactic Chimneys and *Fermi* Bubbles.

*How will AXIS measure it?* Simulations show that an exposure time of 5 ks (250 ks) enables a sensitivity limit of $10^{-15}$ ($10^{-16}$) erg cm$^{-2}$ s$^{-1}$. This would represent 1—2 orders of magnitude below previous surveys in the 0.5—10 keV band despite having equal or less exposure. *AXIS* is ideal for imaging crowded X-ray populations and diffuse emission thanks to its angular resolution and imaging/spectral sensitivity over a wide FoV. The GPS plays to these qualities with a large survey area and a uniform exposure. Once *AXIS* constrains the emission from point sources within its narrow PSF, and has defined their SEDs, the residual diffuse emission can be studied in unprecedented detail.

The GPS addresses several science questions posed by the Astro2020 Decadal Survey [250]:

- What is the population of non-interacting or isolated NSs and stellar-mass BHs? (B-Q1b)
- What powers the diversity of explosive phenomena across the electromagnetic spectrum? (B-Q2)
- Why do some compact objects eject material in nearly light-speed jets, and what is that material made of? (B-Q3)
- What are the endpoints of stellar evolution?
- What are the progenitors and explosion mechanisms of supernovae? (COEP2)
- How do relativistic winds and jets interact with and energize the surrounding medium? (COEP3)
- What are the most extreme stars and stellar populations? (G-Q1)
- How does multiplicity affect the way a star lives and dies? (G-Q2)

Many of these questions can be answered by creating a census of X-ray populations that represent the final stages in the lives of stars. This is especially true for the faint end of each population's luminosity function where *AXIS*, thanks to its low detector background and high spatial resolution, will detect X-ray sources too dim to be seen by other telescopes. In addition to accessing the faint end of the X-ray luminosity function for compact objects, *AXIS* will also allow us to discover younger, high-energy diffuse sources associated with the explosive outcomes of stellar evolution, namely SNRs, PWNe and MWNe. Studying these objects helps us address the question on what happens to stars after they die, sheds light on the explosion mechanisms and progenitors of SNe, and reveals extreme particle accelerators in the Universe. *Chandra* and *XMM-Newton* have opened a new window to linking SNRs to their progenitors, revealing a zoo of neutron stars and associated outflows, and showing pulsar winds as powerful cosmic accelerators of positrons up to PeV energies. However, given *Chandra*'s restricted FoV and its sensitivity degradation, it cannot be used effectively as a survey instrument which essentially results in a bias towards the brighter population. On the other hand the *XMM-Newton* background is too high preventing the detection of faint and extended sources, or limiting our ability to resolve compact Pulsar or Magnetar Wind Nebulae. Therefore, there remains a significant discovery space for *AXIS* in terms of endpoints of stellar evolution and death, which will be enabled by the unique large area and high sensitivity to be achieved with the GPS.

The following sections detail how *AXIS* will help answer these questions.

## 2.1. Ultra Compact White Dwarf Binaries

Type Ia supernovae are important standard candles that arise from the explosion of a white dwarf. There has been significant debate over the years as to the physical mechanism underlying these important explosive transients, with two main progenitor scenarios, the single-degenerate channel in which a white dwarf accretes matter from a non-degenerate object and then explodes, or the double degenerate channel, originating from the interaction of two white dwarfs. The rate of type Ia supernovae in the Milky Way is $\approx 0.3 \times 10^{-2}$ yr$^{-1}$, and it is estimated that the white dwarf merger rate is approximately $4.5 - 7$ times this [224]. When a double white dwarf system starts to undergo Roche-lobe overflow, it can undergo a unique form of accretion known as direct impact accretion, in which the accretion stream directly impacts the surface of the more massive white dwarf rather than forming an accretion disk, resulting in significant X-ray emission, and strong periodic modulation of these X-rays. For merging double white dwarfs with close to equal mass ratio (e.g., a 0.6 $M_\odot$ plus a 0.8 $M_\odot$ white dwarf), this phase is likely short-lived, as the lower mass white dwarf is so dense that it does not undergo Roche lobe overflow until an orbital period of just 40 seconds, at which point the gravitational wave merger timescale is only $\sim 10^2$ yr. If all type Ias originated from this channel, we would only expect $\sim 1$ such accreting system to exist in the Galaxy at any given time. However, recent work has indicated that it is possible that many Type Ia supernovae originating from double degenerate progenitors arise from lower mass helium core white dwarfs accreting onto more massive carbon-oxygen core companions, in many cases via physical mechanisms such as the dynamically driven double degenerate double detonations (D6) [330]. Such systems can undergo Roche-lobe overflow at orbital periods of $>$ 5 minutes, with gravitational wave merger timescales of $\approx 10^5$ yr, suggesting that $\approx 10^3$ such systems might exist in the Galaxy, including dozens which might trigger supernovae in the future. Additionally, many of these systems undergo stable mass transfer and become AM CVn-type binaries, and recent work with the Zwicky Transient facility has demonstrated the existence of many new eclipsing AM CVns with ultrashort orbital periods (Burdge et al., in prep). With the 104 square-degree *AXIS* Galactic plane survey at a depth of $\sim 10^{-15}$ ergs$^{-1}$, we expect to be sensitive to disk-dominated systems such as the recently discovered ZTF J0127+5242 at a distance of $\sim$ 8 kpc, allowing us to probe approximately $\sim 5$ percent of such systems in the Galaxy. We include a portion of the relatively unobscured Bulge region in our footprint to maintain sensitivity to direct impact accreting systems such as HM Cancri, which have very soft spectra and are easily obscured. We estimate that at a column density of $N_\mathrm{H} = 2 \times 10^{21}$ cm$^{-2}$, we would be able to blindly detect a source like HM Cancri in periodicity search at a distance of $>$ 20kpc in a 6 ks *AXIS* exposure. Current estimates indicate that $2 \times 10^{10}$ $M_\odot$ of stellar mass resides in the Galactic bulge [363], and $5.17 \times 10^{10}$ $M_\odot$ in the disk [209], indicating that the Bulge hosts approximately 28 percent of the stars in the Galaxy. These stars are distributed over approximately 380 square degrees, which means the stellar density is high enough such that each square degree contains approximately 0.073 percent of stars in the Galaxy. Thus, by including 78 square degrees of relatively unobscured regions of the Bulge in the *AXIS* GPS, we will systematically probe around 5.7 percent of stars in the Milky Way for direct impact accretors (after accounting for only half the sources being eclipsing, this should be about 3 percent of direct impact systems in the Galaxy). Thus, if there are 1000 direct impact accretors in the Galaxy, we expect $60.00 \pm 7.75$ direct impact systems to reside in this 80 sq degree footprint in the Bulge. Assuming we can confidently detect half of these systems (since approximately half should undergo full X-ray eclipses), if we do not detect any systems and they trace the stellar mass of the Galaxy, we will have ruled out there being more than 320 such systems in the Galaxy with 3$\sigma$ confidence (e.g if there are 320 such systems in the Galaxy evenly distributed among the stellar mass, the probability of our search identifying zero of them given that half are eclipsing is 0.0003). Given the rate of inspiraling double degenerates, we estimate that in our full survey footprint we will be able to detect approximately 30 pre-period minimum mass-transferring ultracompact binaries including direct impact systems and disk-dominated systems such as ZTF J0127+5242 (and an additional 30 eclipsing post-period minimum AM CVn objects such as the recently discovered 55-minute eclipsing AM CVn in the *SRG* eFEDs field). We plan to leverage the five-year baseline of the survey to measure orbital evolution in any detected systems to determine whether they are inspiraling and are likely to merge/trigger a supernova, or are undergoing stable mass transfer and likely to evolve into an AM CVn type system.

Our efforts are inspired by the shortest orbital period binary known in the Galaxy, HM Cancri (orbital period 5.35 minutes), which is a direct impact accreting system that was discovered using the *ROSAT* all-sky survey because it strongly modulates its X-ray flux on its orbital period. In this remarkable system, the hot spot on the accreting white dwarf is eclipsed every orbit of the binary, leading to a 100 percent modulation of the X-ray flux, as shown in Figure 5. Long-term timing of the X-ray waveform has also revealed strong orbital decay due to the emission of gravitational radiation, and remarkably, has led to a measurement of the second derivative of the orbital frequency, which indicates that the mass transfer in the binary is counteracting the angular momentum loss due to GR and is slowing down the decay rate and that the system should reach a minimum orbital period in just a few thousand years. One can use the masses inferred from this orbital evolution to conclude that HM Cancri is unlikely to be a type Ia progenitor, and will instead likely transition to stable mass transfer. This source illustrates how a sensitive X-ray time domain survey can be used both to identify ultracompact binaries, and via timing measurements, characterize the chirp mass of the system, and thus whether it is likely to be a Type Ia supernova progenitor. By using *AXIS* to conduct a survey of the Galactic plane, we will leverage its sharp PSF over a wide field of view to create the highest spatial resolution X-ray map of the Galactic plane, and its effective area will allow us to collect enough spatially resolved photons with cadenced temporal resolution so that we can use periodicity analyses on these X-ray sources to discover sixty new interacting ultracompact binaries detectable in the LISA band. Using the five-year baseline of the survey, we will measure orbital evolution in these systems (as seen in panel b of Figure 5) to determine whether they are likely to result in a type Ia supernova. By discovering 60 such systems, we can confirm or rule out whether more than a few percent of the interacting double degenerate population in our Galaxy are likely to result in Type Ia supernovae. Additionally, by measuring which systems are undergoing stable mass transfer and evolving out to longer orbital periods, we may be able to constrain which combinations of white dwarf masses lead to merger and explosion, and which lead to stable mass transfer.

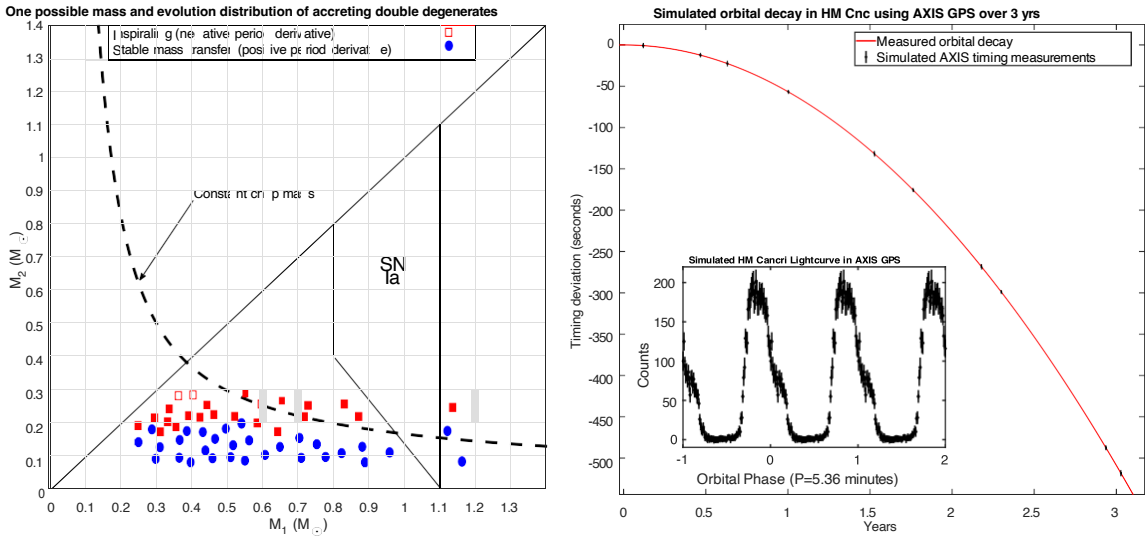

**Figure 5.** Left: White dwarf mass distribution of one possible mock population of detected accreting ultracompact binaries in the *AXIS* GPS, with blue circles indicating stably mass transferring sources, and red squares indicating inspiraling sources. The background of this plot indicates one hypothetical region of mass vs mass space where Type Ia supernovae are likely to occur. Right: Simulated timing measurements of the X-ray waveform of HM Cancri over three years of the *AXIS* GPS, illustrating our sensitivity to detecting the period derivative of such sources in the multi-epoch longer baseline survey. Inset: A simulated lightcurve of HM Cancri at a distance of 8 kpc given the 6-ks exposure of the *AXIS* GPS. The periodic X-ray modulation in this source is significant enough that it is easily recoverable in a blind periodicity search of a million sources.

*2.2. Population Studies from the proposed GPS*

### 2.2.1. Machine Learning Classification

With *AXIS*'s sufficient sensitivity, large FoV, and excellent PSF, we will be able to detect many more (probably millions of) X-ray sources with arcsecond localizations, from the surveys like the GPS, compared to previous and existing observatories (e.g., *Chandra*). In such an era of large data astronomy, rapid classification of a huge number of sources becomes a particularly important task. New software and methods have been developed, to enable cross-matching sources from different surveys and catalogs [322], and to apply on-the-fly classification of X-ray sources using methods like machine learning (ML) [357,399]. Among those, a multiwavelength machine learning pipeline for classification of X-ray sources (MUWCLASS) [399] has been developed by colleagues from our CO/SNR science group.

The major component of a supervised ML pipeline is the training dataset (TD), which is a collection of sources with confident classifications. In MUWCLASS, there are several thousands of X-ray sources in the TD which are categorized into 8 classes of X-ray emitters including active galactic nuclei (AGNs), cataclysmic variables, high-mass stars, HMXBs, low-mass stars, LMXBs, NSs, and young stellar objects.

*AXIS* will accurately measure the X-ray energy fluxes at multiple energy bands, the X-ray hardness ratios, and the X-ray variability properties of each X-ray sources, given its high sensitivity and the relatively good timing resolution (of $\sim 0.2$ s or even better). Besides the X-ray features, the photometric properties (e.g., magnitudes and colors) at lower frequencies are also crucial to ML classifications to break the degeneracy of classifications when using only the X-ray properties. The sub-arcsecond localizations from *AXIS* observations make it possible to enable accurate cross-correlating between the X-ray sources and multiwavelength surveys at lower frequency, especially in a crowded environment like the GPS.

MUWCLASS is trained with a supervised ensemble decision-tree algorithm called random forest. To mitigate the bias of the extinction/absorption of AGNs from the TD which come from surveys conducted away from the Galactic plane, MUWCLASS applies location-specific reddening/absorption corrections to AGNs from the TD while classifying sources in the Galactic plane. It also uses an implementation of the synthetic minority over-sampling technique to oversample the TD [58] to overcome the large imbalance of source types from the TD. Measurement uncertainties are also taken into account by Monte Carlo (MC) sampling from feature probability density functions and averaging multiple MC sampling results to obtain confident classifications and measure their uncertainties. The performance of MUWCLASS has an overall accuracy of about 86%, up to 95% for confident classifications.

Moderate-resolution CCD X-ray spectra, by themselves, can also be used to distinguish different kinds of X-ray sources. Chromospherically active stars emit strong Fe-L and Ne-K lines ($\sim 1$ keV), and SNRs have dominant Mg-K ($\sim 1.3$ keV) and Si-K ($\sim 1.8$ keV) line emissions. The X-ray spectra of NS and AGN are not line-dominated (though AGN can also have a fluorescent Fe line at rest energy of 6.4 keV). Thus, ML algorithms that can distinguish between line-dominated and continuum-dominated sources can separate X-ray sources from their X-ray spectra alone. Such methods can either be used as stand-alone classification algorithms when multi-wavelength observations and cross-matching are not feasible, or in combination with other multi-wavelength classification algorithms. Hebbar & Heinke [145] tested this idea by using an artificial neural network (ANN) to differentiate the *Chandra* spectra of stars in the *Chandra* Orion Ultradeep Project (COUP) from the spectra of AGN in the *Chandra* Deep Field South (CDFS) catalog and separated the two classes with an overall accuracy of 90%. Their results found that the accuracy of the ANN classification is greater than 90% for sources with net counts greater than 200 and background contribution less than 5%. For a 100 ks *Chandra* exposure, this implies that we can accurately ($> 90\%$ confidence) classify sources down to a flux of $5 \times 10^{-14}$ ergs cm$^{-2}$ s$^{-1}$ in the 0.5–10 keV energy range. Sources with high absorption ($N_{\mathrm{H}} > 10^{22}$ cm$^{-2}$) are not properly classified indicating that detecting soft energy photons < 2 keV is crucial to distinguish between Poisson noise and emission lines. The degrading soft energy response of *Chandra* ACIS makes it difficult to combine observations from different epochs and reduces the number of soft energy X-ray photons detected, thus negatively affecting the classification accuracy.

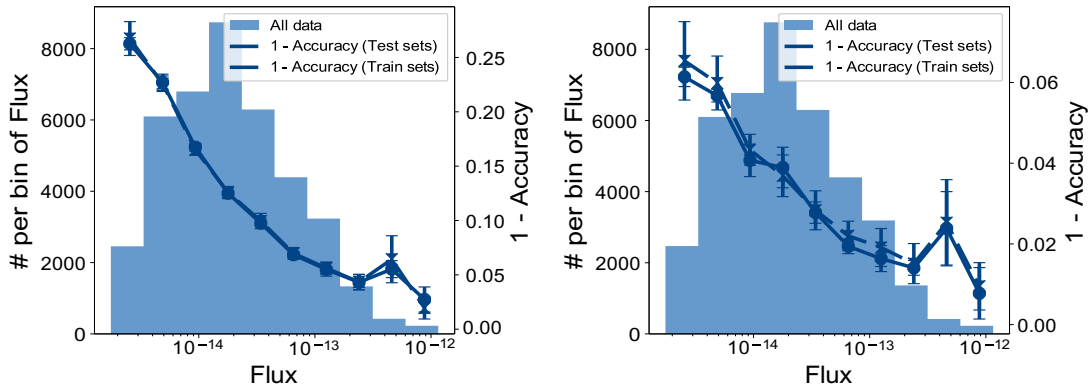

**Figure 6.** Comparison of ML classification on *Chandra, (Left)* and *AXIS* (*Right*) spectra with exposure of 100 ks using a 1-layer, 10-node ANN. The *x*-axis shows the absorbed flux in the 0.3–8.0 keV energy range. The left *y*-axis shows the number of simulated spectra in each bin of the histogram, and right *y*-axis shows the error in classification with flux. The dashed lines show the error in the *training* dataset used for fitting the ANN, and the solid line shows the error in applying the trained ANN to an independent test set. We see that with *AXIS* we can accurately classify >90% of sources throughout our range of fluxes used for simulation. Note that we still require $N_H < 10^{22}\,\mathrm{cm}^{-2}$ and >200 net counts for identifying line-dominated star spectra, but the larger effective area and better soft energy response allows us to probe fainter sources.

*AXIS* can provide significant improvements in this area due to its larger effective area and field of view, reduced detector background, and better soft energy response. We use the distribution of COUP and CDFS parameters with *AXIS* FoV-average effective area, response-matrix files, and non-X-ray background, and 100 ks exposure to simulate X-ray spectra of stars and AGN with *AXIS* and test the improvements in classification accuracy. Our results show that we can classify active stars and AGN with an accuracy of > 90% throughout our range of absorbed flux i.e > $2 \times 10^{-15}$ ergs cm$^{-2}$ s$^{-1}$. In terms of net counts, we still need > 200 net counts and $N_H < 10^{22}\,\mathrm{cm}^{-2}$ for classifying with an accuracy of > 90%. If we assume that the *AXIS* effective area is 10 times that of *Chandra* and assume a similar exposure map of *Chandra* and *AXIS*, we can estimate that sources with >= 20 net counts in the *Chandra* ACIS 0.5-7 keV can be classified accurately. The *Chandra* Source Catalog 2.0 has ~ $2.1 \times 10^5$ such sources. Among these sources, $1.7 \times 10^5$ sources have Galactic $N_H < 10^{22}\,\mathrm{cm}^{-2}$. In comparison, $3.4 \times 10^4$ sources have net counts > 100 ($2.5 \times 10^4$ of these source also have Galactic $N_H < 10^{22}\,\mathrm{cm}^{-2}$). Therefore, incorporating *AXIS*'s 9 times larger field of view, we can anticipate that *AXIS* should be able to classify of order of 35 to 60 times more X-ray sources than can be classified using *Chandra* data.

### 2.2.2. Millisecond pulsar population in Galactic Bulge

The *Fermi*-LAT observations of the Galactic center show an excess in the γ-ray emission (Galactic Center Excess: GCE) in addition to that from known point sources and the Galactic ridge. While some studies have shown that this excess radiation is consistent with annihilating dark matter [e.g., 85,160], others have suggested that the GCE could be explained in its entirety through a population of unresolved millisecond pulsars (MSPs) [e.g., 2,114]. As MSPs also emit X-rays, we will be able to detect individual MSPs through X-ray telescopes having much better angular resolution than *Fermi* (*Chandra* has an aimpoint resolution of 0.5″, and *XMM-Newton* has a resolution of 10″). Thus understanding the population of the MSPs in the Galactic center and bulge will allow us to put further constraints on the origin of the GCE. However, the soft energy X-rays (below 2 keV) from most MSPs in the Galactic center and bulge are absorbed by the high interstellar extinction towards that region. Only bright MSPs with hard spectra (through magnetospheric X-ray emission or from intra-binary shock in spider MSPs) can be detected by X-ray telescopes. If all of the GCE is from an MSP population with properties similar to that of globular clusters, we will be able to detect 1–86 MSPs with $L_{X,0.3-8keV}$ > $10^{33}$ ergs s$^{-1}$ or 20–910 MSPs with $L_{X,0.3-8keV}$ > $10^{32}$ ergs s$^{-1}$ in the Galactic bulge [410].

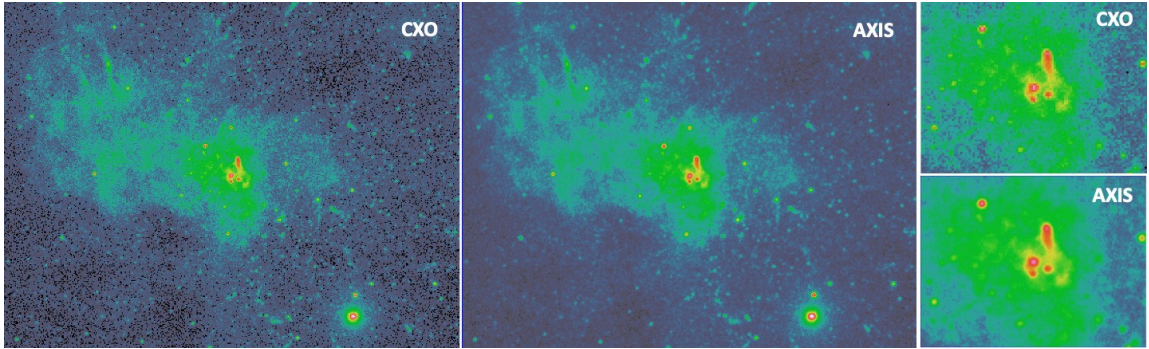

**Figure 7.** *Left:* 1 Ms of *Chandra*/ACIS-I imaging of the GC obtained during its first 10 years in orbit (∼ 2.0′ × 1.45′). *Right:* Simulated *AXIS* image of the same region as depicted on the left. The summed exposure time is 102 ks, consisting of 34 × 3 ks exposures randomly dithered within a 1′ × 1′ box centered on Sgr A*. The subsequent stack is combined using drizzle methodology and enables statistically significant detection of all sources in the ACIS-I image. The cutouts on the right show zoom-ins on Sgr A* (∼ 27″ × 20″).

Hard X-ray emission from intermediate polars (IPs) and background AGN can cause confusion in the type of source detected. Spectroscopic analysis or identifying Fe lines might be crucial to distinguish MSPs and IPs. *XMM-Newton* observations of the Galactic bulge suffer from a large amount of diffuse emission in the region and we can only distinguish whether Fe lines are present or not in point sources with luminosities down to $L_{X,0.3-8keV}$ = $2 \times 10^{33}$ ergs s$^{-1}$. While *Chandra* has a smaller PSF, and thus less contribution from the diffuse emission, it would need tens of Ms to scan the Galactic bulge at sufficient depth to find ∼10 MSPs. The larger effective area, wider FoV, and sub-arcsecond resolution of *AXIS* should allow identification of large numbers of MSPs with the proposed GPS.

## 3. Galactic Center and Sgr A*

The Milky Way is an archetypal disk galaxy, lying near the peak of the galaxy mass function. The Milky Way contains a SMBH at its center that lies on the $M_{BH}$–σ relation. At a distance of 8 kpc (1″ ∼ 0.04 pc), our Galactic nucleus presents the observational test-bed in which to understand the extremes of star formation and black hole growth in our home galaxy. Deep spatially uniform imaging spectroscopy of the Galactic center will place direct constraints on feedback processes in an average disk galaxy as the outflows (kinetic/radiative) from Sgr A* and the vigorous star-formation taking place in the Galactic center interact with the immediate environment, the nuclear star cluster, nuclear stellar disk and the reservoir of molecular gas in the CMZ. This program of observations will inform modeling efforts to understand the physical processes driving the observed connection between galactic bulges and their super massive black holes. All of this science requires a next generation high spatial resolution X-ray observatory – *AXIS*. An example of the expected performance of *AXIS* for Galactic center science is displayed in Figure 7.

Sgr A* is the supermassive mass black hole at the center of the Milky Way, $M_{BH}$ = (4.07 ± 0.1) × $10^6$ $M_\odot$, $d_{GC}$ = 8.1 ± 0.1 kpc [110,242]. Pioneering observations with *Chandra* have revealed X-ray emission from the accreting black hole, as well as emission from the broader Bondi-flow feeding the black hole [23,24,382]. Sgr A* is a low luminosity black hole with an Eddington scaled accretion rate $L_x/L_{Edd} \sim 10^{-9}$ erg s$^{-1}$($L_x \sim 10^{33}$ erg s$^{-1}$, 2-10 keV). Studies of our black hole thus provide insight into the canonical accretion mode of supermassive black holes in the Universe.

### 3.1. Sgr A* Flares

Sgr A* is known to have a moderate $\sim 10\times$ flare at X-ray energies on a daily basis [251,252]. These flares are accompanied by counterparts in the nIR and sub-mm/radio [88,89]. Current campaigns have presented tentative evidence for the nature of the flaring mechanism; however, the difficulty of planning large multi-wavelength campaigns places limits on the observing opportunities given the known flaring rate of Sgr A* [251,252]. *AXIS* will provide advances in 2 primary ways, (i) the increased sensitivity will open up a population of more common lower luminosity flares to detailed study and, (ii) the scheduling flexibility of the observatory will aid the planning of the required large multi-facility campaigns.

The study of flaring from Sgr A* demands a multi-wavelength approach and observations across the electromagnetic spectrum are required to extract constraints on the physical mechanism responsible for the flaring, e.g., [42,43,95,388]. The mechanism driving the variability observed across the EM spectrum has not been identified, with a number of competing models proposed, ranging from magnetic re-connection or instabilities in the accretion flow, to jetted ejection events and expanding plasma "blobs" [83,86,87,90, 208,226,403,406].

The Sgr A* flare X-ray flux distribution has been constrained by *Chandra*. The quiescent emission is consistent with a steady Poisson process component ($f_x \sim 4.5 \times 10^{-13}$ erg s$^{-1}$ cm$^{-2}$, $2 - 8$ keV) in addition to flaring which can be described with a powerlaw ($\xi \sim 1.9$). When characterized as a log-normal process, the median flare flux is $\sim 4 \times 10^{-14}$ erg s$^{-1}$ cm$^{-2}$ ($2 - 8$ keV), with the flaring component contributing 10% – 15% of the quiescent flux [251,252]. *AXIS* observations will revolutionize the study of Sgr A* flaring activity, facilitating constraints on the X-ray spectra and morphology for flares deep into the known flare luminosity function. The most luminous flares ($\gtrsim 1000\times$) are rare but provide some of the best constraints on the broadband emission mechanism [136,259,287]. *AXIS*' observing modes will enable high time resolution spectroscopy during these events that will compliment the advanced capabilities of multi-wavelength facilities in the 2030s.

The Gravity and EHT projects have provided pioneering advancements in our capability to study Sgr A* in recent years [93,94,121,122,353]. These projects have plans to enhance their capabilities for the 2030s [123,173]. A sensitive high spatial resolution X-ray observatory is required. *AXIS*' combination of scheduling flexibility, flux sensitivity and spatial resolution will ensure access to the physics of these flares at X-ray energies. These X-ray observations will complement coordinated multi-wavelength observations from facilities such as ELT, Roman, Gravity+, and ngEHT promising a revolution in our understanding of the accretion flow, relativistic spacetime, and the SMBH at the center of our galaxy.

An opportunity of particular interest will present itself in 2034 when S2 will make its next periastron passage $\sim 120$ AU ($\sim 2800\ r_s$) from Sgr A*. Observations in the years surrounding the 2018 periastron passage have presented evidence for a changing flaring rate temporally associated with this event and the periastron passage of the G2 object [16,243,286]. Coordinated observations across the EM spectrum are required to constrain the flaring mechanism, and the flexible high spatial resolution X-ray observations provided by *AXIS* will be the only way to constrain the X-ray emission during these campaigns in the 2030s.

### 3.2. The Quiescent Accretion Flow & Bondi Flow

The quiescent emission from Sgr A* is consistent with thermal bremsstrahlung in the X-ray band, emitted by an outflow-dominated accretion flow within the BH Bondi radius [382]. Observations with *AXIS* will open up the possibility of new time domain studies of this quiescent accretion flow. *Chandra* observations have demonstrated the flow to be outflow dominated. *AXIS* observations will probe the variation of this plasma for the first time and constrain the variation of the continuum with respect to variations in the Fe XXV emission line revealed by *Chandra*.

Of particular interest are the interactions of the S-stars with the hot gas, e.g., S2 approaches ∼ 120 AU (∼ 17 light-hr) from Sgr A* during periastron passages [121]. The next passage will occur in 2034 and observations of the quiescent flow and its response to the S2 star passage will be illuminating. There are also an emerging population of so-called G-objects in the vicinity of the Sgr A* in addition to a recently identified filamentary feature that may be breaking up as it approaches the SMBH, e.g., [66].

*AXIS*' sensitivity will enable studies of the spatial morphology of the diffuse emission associated with the accretion flow in addition to placing constraints on the radial temperature distribution of the plasma. Winds from massive stars in the central 2 pc have been shown to be capable of fueling the SMBH [48,74,75,303,303,339], and these winds in turn can interact with and disrupt the circumnuclear disk, driving further gas towards the black hole [303]. Of key interest will be going below the sensitivity limit of *Chandra* to study the extended diffuse emission in the outer accretion flow and place observational constraints on hot plasma in the transition region where the stellar winds from stars in the circumnuclear disk feeding the SMBH slow and enter the black hole's sphere of influence [74,75,339]. Potential variable jetted emission from Sgr A* may also impact the inflowing gas [74,75,339].

### 3.3. Stellar Populations in the Galactic Center

Stellar evolution predicts a large population of stellar mass BHs and neutron stars in the Galaxy. In the Galactic center, dynamical friction should result in the accumulation of a large number of stellar mass BHs in the vicinity of Sgr A* [100,231,236,241]. Many X-ray sources have been detected in the GC by *Chandra* [159,244,245,247,415] and a fraction of these have been interpreted to display the characteristics of a population of quiescent stellar mass black holes [137,240]. The nature of these sources is contentious [218,239] and high spatial resolution X-ray observations are required to enable further progress on this question.

*AXIS* observations will also facilitate observations of massive star forming clusters (e.g., Arches, Quintuplet, IRS 13E), those in the process of forming (e.g., Sgr B2), in addition to X-ray sources in the nuclear star cluster and nuclear stellar disk [51,101,105,111,379,381].

### 3.4. Diffuse X-ray Emission

*AXIS* will carry out a deep uniform survey of the Galactic center. The combination of excellent sensitivity in the 2–10 keV bandpass and a uniform PSF across the FoV will enable the detection of the hot plasma content extending from Sgr A* to the circumnuclear ring (∼2 pc, ∼ 50″) and out to the CMZ (r ∼200 pc, ∼ 80′) and beyond. Spatially uniform *AXIS* imaging across the Galactic center will reveal the X-ray emission morphology in this entire region at arcsecond resolution (∼0.04 pc at 8 kpc), and enable constraints to be placed on the physical state of the hot plasma and the source populations/processes which generate it, e.g., [15,203,245,257,360].

A key *AXIS* goal will be searching for connections linking the abundant plasma filling the GC and the *Fermi* bubbles [348] (with prominent counterparts observed at X-ray and radio wavelengths [54,297]). Observations on smaller spatial scales have demonstrated the existence of the so-called GC chimneys, which reveal 100-pc scale X-ray outflows from the GC region towards the *Fermi* bubbles [249,288]. The observation of molecular gas counterparts demonstrates the multi-phase nature of this outflow, and presents a prime example of ongoing in-situ feedback [284,372]. In the immediate vicinity of Sgr A*, X-ray emission suggesting the presence of jets originating from the black hole will be constrained [407,416].

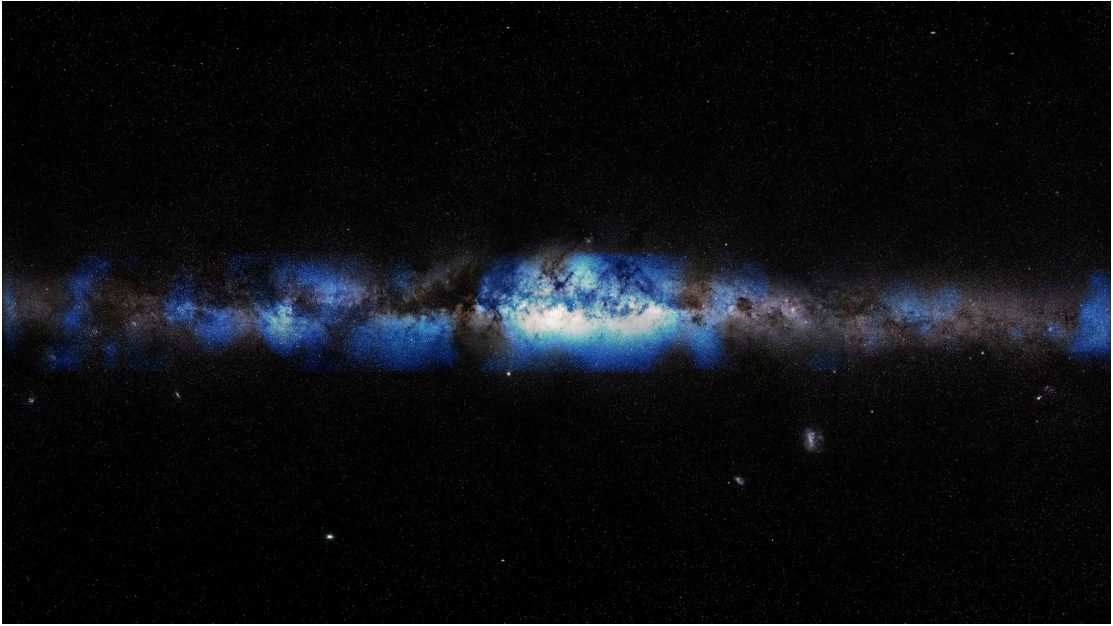

**Figure 8.** An artist's composition of the Milky Way seen with a neutrino lens (blue). Credit: IceCube Collaboration/U.S. National Science Foundation (Lily Le & Shawn Johnson)/ESO (S. Brunier).

Large scale mapping of the GC at radio wavelengths has revealed diverse extended structures (see Fig. 2, [152]). The thin filamentary structures detected therein, unique to the GC, are of particular interest and are thought to be a result of the intense SNe and SMBH feedback in the GC [338,405]. Molecular clouds near Sgr A* have been shown to exhibit strong and variable X-ray fluorescence [191,285]. *AXIS* observations will constrain the location and morphology of the Fe K-shell emitting gas on unprecedented spatial scales, enabling study of the spatial distribution and structure of dense molecular gas in the Galactic center [65]. Comparison of Fe-K maps on multi-year timescales will enable a sensitive study of its variation across the molecular clouds with time. These observations will facilitate the creation of a high time resolution record of the past activity of Sgr A* [52,64,68,225] and provide constraints on the cosmic ray population [310] in the Galactic center.

The proposed *AXIS* studies of the GC region will have a clear multi-messenger component given the *IceCube* detection of a neutrino flux consistent with an origin in the Galactic plane ([168], see Fig. 8). Searches for a discrete point-like (e.g., XRBs, pulsars, Sgr A*) or extended (e.g., SNRs) source population of Galactic neutrino emitters has thus far returned null-results, although a number of candidate sources are present at $< 3\sigma$ level [3,4]. As the S/N continues to build in coming years we can expect a catalogue of Galactic neutrino sources to emerge. This discovery opens a new path to study the Milky Way and the high spatial resolution and sensitivity of *AXIS* will play a key role in our exploration of the sites of Galactic cosmic-ray generation and the multi-messenger Milky Way in the next decade.

## 4. Clusters (Open and Globular)

Open clusters are the primary birthplaces of compact objects (COs; referring here to neutron stars and black holes) in the galaxy, but given that for most of them, their total masses are relatively low ($< 1000 \ M_\odot$ [278,279]) and not centrally concentrated (i.e., as in globular clusters), a majority of COs quickly escape their birth clusters (see e.g., [235,329,367]). However, COs formed in very massive ($M_\odot > 10^4$) and young ($\lesssim 10$ Myr) stellar clusters may still reside in them. In this respect, the *Chandra* discovery of a young magnetar in the outskirts of the $\sim$ 5-Myr-old Westerlund 1 open cluster is encouraging [246]. However, if other, less luminous COs (e.g., older magnetars, NSs, or BHs with a low level of accretion), exist in these clusters, they would be difficult to find/identify with the existing X-ray observatories. Discovering such objects in their natal environments allows one to place very tight constraints on the properties of the progenitor star and SN explosion models [246]. *AXIS* will enable the discovery of new COs in open clusters given its high angular-resolution (necessary to resolve the many point sources in crowded environments and to provide accurate source positions), high sensitivity, and large field of view. As open clusters age, the most massive of them remain bound, retaining a fraction of their stars. It's possible that these older clusters can retain COs, if a fraction of the COs experience small kicks and/or are born in binaries. For example, neutron stars born from electron-capture supernovae are thought to receive low enough kicks to be still bound to their birth clusters [112,169,344]. *Chandra* studies of open clusters, with varying ages, have uncovered a plethora of different source types, but the only CO confirmed thus far is the magnetar in Westerlund 1 (see e.g., [60,113,364,371]).

Globular clusters (GCs) host an overabundance of accreting COs per unit mass when compared to the Galactic disk. This is due to the large number of dynamical interactions occurring in the dense cores of these clusters, which leads to an increased number of tight binaries [67]. Due to extreme crowding in the cores of GCs, an X-ray observatory with sub-arcsecond resolution is needed to study the cluster's X-ray source populations. *Chandra* has played a pivotal role in advancing our understanding of X-ray source populations in GCs, particularly at the fainter luminosity end (see e.g., [135,147–149,151,290]). A number of interesting source classes hosting COs are typically found at these fainter luminosities (i.e., $L_X \lesssim 10^{32}$ erg s$^{-1}$), including quiescent low-mass X-ray binaries with NSs (qLMXBs) [150], MSPs (see e.g.,[35,36,410]), and an emerging population of candidate BH LMXBs [25,63,234,345], one of which recently underwent an outburst and produced radio jets [26,277]. Discovering more of these systems allows for understanding the dynamical production and destruction of binaries, as a function of encounter rate, metallicity, cluster structure, and other properties (see e.g., [289,291]). *AXIS* will enable deep and efficient surveys of a majority of the Galactic GCs down to luminosities of $L_X < 10^{30}$ erg s$^{-1}$. Pushing to lower luminosities in a large number of GCs is critical for identifying new unique objects, such as BH LMXBs, several of which have quiescent luminosities $L_X < 10^{30}$ erg s$^{-1}$ [144,345]. By identifying more of these systems (in concert with deep radio studies), we can start to understand their formation rate in comparison to the binaries hosting NSs. This has important implications for the rate of BH-BH mergers being detected by LIGO (see e.g., [17,18,409]).

GC science is also inherently a multiwavelength science. *AXIS* will be able to leverage the rich pre-existing multiwavelength data sets from radio to GeV wavelengths, as well as new observations from, for example, *JWST*. Optical and IR data from *HST*, *JWST*, VLT/MUSE, Gemini/GIRMOS, and *Gaia* can be used to place tight constraints on the distance, age, mass, and metallicity of GCs, and to search for the optical/IR counterparts of the X-ray emitting sources (see Figure 9 and e.g., [142,148,326]). Having very accurate distance measurements allows for more accurate estimates of NS masses and radii (see e.g., [37,56,130]), which are necessary for constraining the NS equation of state. Additionally, the physics of the crust of NSs can be probed by observing how they cool after undergoing an accretion episode (see [80,262,385]). Another example is the MSP population of GCs, which is typically uncovered by radio observatories and then can be followed up in X-rays (see e.g., Terzan 5 [35,299]). In particular, X-ray observations have shown that this population is diverse ranging from isolated MSPs to tight spider binaries. These systems are composed of redback binaries, where the MSP wind interacts with the low-mass companion, and the even more extreme black-widow binaries, where the pulsar wind completely ablates the companion star (see e.g., [346,350]). A repeating Fast Radio Burst (FRB) has also recently been discovered in an extragalactic GC [195], suggesting GCs host some fraction of repeating FRBs. Sensitive X-ray observations of Galactic GCs will help to test FRB models and what types of sources are capable of producing them. As new observatories come online over the next decade (e.g., SKA, CTA, Roman) and discover new COs, the lack of a next generation sensitive, high angular resolution X-ray observatory to replace *Chandra*, would greatly hinder Galactic GC science (see Figure 9). *AXIS* will enable the continuation of this science and inevitably lead to many new discoveries that will help to further our understanding of COs and GCs.

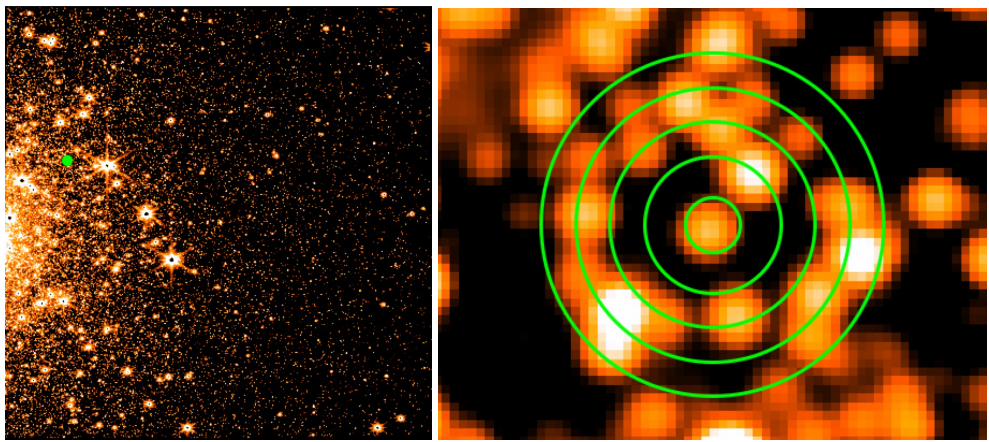

**Figure 9.** *Left: JWST* NIRcam F277W image of the Galactic globular cluster M92. The green point shows the location of the zoom in on the right panel, note that it is outside of the core of the cluster. *Right:* Zoom-in showing hypothetical positional error circles of an X-ray source outside of the cluster core. The circles have radii of 1.25", 1", 0.75", 0.5", and 0.2", respectively. Note that the number of potential IR counterparts detected by *JWST* drops dramatically with increasing X-ray positional accuracy, afforded by *AXIS*'s exceptional PSF, allowing for confident identifications of optical/IR counterparts to X-ray sources.

## 5. Compact Objects: Isolated Neutron Stars

A neutron star is the left-over core of a massive star that underwent a supernova explosion. When the progenitor is massive enough, the star may collapse to compress the core to the density at which neutron degeneracy pressure kicks in and keeps the star from further collapse. The density of the core at this point is above nuclear density, and the core has a radius of $\sim 10$ km and a mass of $\sim 1.4 M_\odot$. A neutron star can have a rapid spin and strong magnetic field ($B$), e.g., if the angular momentum and magnetic flux of the progenitor were conserved during its collapse. Neutron stars provide a unique laboratory to probe the most dense matter and the highest $B$ in the Universe, giving opportunities to revolutionize our understanding of physics in extreme environments [see 91,139,185,314, for reviews].

### 5.1. Population of Isolated Neutron Stars

Neutron stars are generally discovered as pulsating (rotating) sources in the radio, X-ray, and gamma-ray band; these pulsating neutron stars are called pulsars. Radio, X-ray, and gamma-ray observatories have served as excellent tools in studying neutron stars for decades, and they helped reveal their observational properties. At the same time, discoveries of the different types or a zoo of neutron stars have complicated the simple and fundamental picture for neutron stars, dense matter under strong $B$, as is displayed in the diagrams (Figure 10) of the spin period ($P$) *vs* its time derivative ($\dot{P}$), and $P$ *vs* the spin-inferred surface dipole magnetic field strength ($B_s = 3.2 \times 10^{19} \sqrt{P\dot{P}}$). Neutron stars have been categorized into several classes: rotation-powered pulsars (RPPs), X-ray dim isolated neutron stars (XDINSs), central compact objects (CCOs), millisecond pulsars (MSPs) and magnetars based on their emission properties and grouping in the $P$-$\dot{P}$ diagram. It was suggested that these neutron-star classes are all linked to one another via magneto-thermal evolution [e.g., 275,295,373], which may help unify the diverse classes into one, allowing us to understand the state of the most dense matter under extreme $B$. We describe basic observational properties of the neutron-star classes in relation to the *AXIS* observatory, and discuss how it can help understand neutron stars better. Here we focus on 'isolated' neutron stars (i.e., not in a binary) and their observational properties.

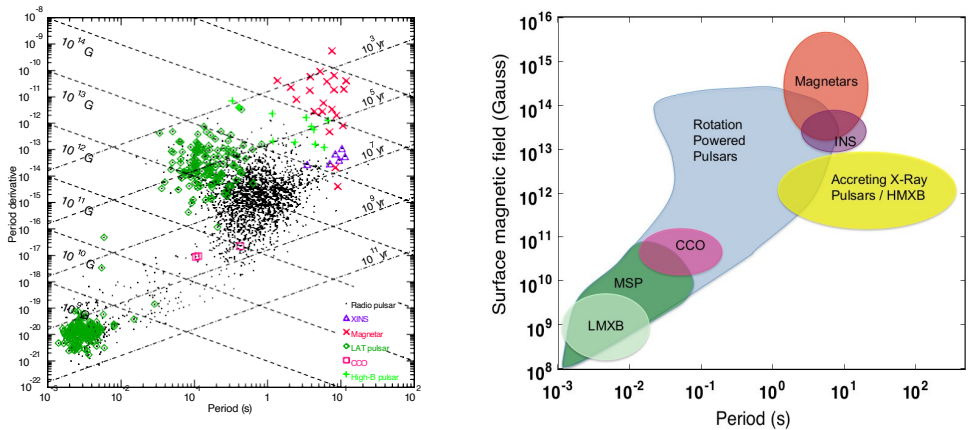

**Figure 10.** (Left): The $P$-$\dot{P}$ diagram constructed using pulsars in the ATNF catalog. (Right): A schematic diagram of $P$ *vs* $B_s$ taken from Harding [139].

**RPPs** are neutron stars whose electromagnetic emission is powered by rotational energy of the pulsar (spin-down power $\dot{E}_{SD}$). They are mostly detected in the radio band, but some of them are detected in the X-ray and gamma-ray bands. X-ray emission from many RPPs is dominated by nonthermal emission, both magnetospheric and from surrounding PWNe. *AXIS* can help distinguish these, via spatial resolution for many PWNe (see Section 7) and, for young pulsars, by gating out pulsed magnetospheric flux. Many RPP sources emit thermal blackbody (BB) emission as well [e.g., 295,414] and some RPPs show only thermal emission. These thermal emissions are particularly important since they can provide important clues to the evolution of neutron stars and fundamental physics; e.g., the low thermal luminosity ($L_{BB}$) of the young PSR J0205+6449 in 3C 58 [337] has been interpreted as due to rapid cooling via the direct URCA process enabled by high proton fractions in the core of neutron stars with mass $\geq 1.6 M_\odot$ for certain nuclear equations of state [295].

**XDINSs** are neutron stars which do not have an associated supernova remnant, a binary companion, or a pulsating radio counterpart. Until now, twelve sources, including candidates, were discovered, and their emission is primarily in the X-ray band. Spectra of XDINSs are well characterized by $kT =$ 0.045–0.1 keV BB emission with broad absorption features (and some UV and X-ray excess). XDINSs have $P = 3 - 17$ s and $B_s \sim 10^{13}$ G which overlap with magnetars in the $P$-$\dot{P}$ diagram (Figure 10). Large $L_{BB}/\dot{E}_{SD}$ values estimated for XDINSs, sometimes exceeding 1 [1RX J0720.4−3125; 178], may imply that their high $B$ plays an important role in their emission like in the case of magnetars. Hence it has been suggested that XDINSs are descendants of magnetars [e.g., 172,185,293] since the former have very similar spin and emission properties to the latter but much larger characteristic age $\tau_c$ of typically $\geq 10^6$ yrs.

**CCOs** are young and radio-quiet isolated neutron stars discovered near the center of SNRs, shining mostly in X-rays and lacking the PWNe expected to be found around young neutron stars. They are typified by the CCO discovered with the first light *Chandra* observation of Cas A [273,351]. Spectra of CCOs are well characterized by thermal blackbody-like emission with X-ray luminosity of $\sim 10^{33}$ erg s$^{-1}$, and one source has strong harmonically-spaced absorption lines [see 78, for a review]. While their spectral properties are similar to those of XDINSs and some quiescent magnetars, CCOs have different spin properties from these neutron star classes [but see properties of the XDINS 1RXS J141256.0+792204 (Calvera); 38,230,408]. Pulsations of CCOs with $P \approx 100 - 400$ ms were detected in only three sources to date [116], and their spin-inferred $B_s$ and $\tau_c$ are $3 \times 10^{10} - 10^{11}$ G and $> 10^8$ yr, respectively (Figure 10). Their relatively low inferred magnetic field led to their classification as 'anti-magnetars' [116]. Another interesting property if their $\tau_c$ values being orders of magnitude larger than the estimated ages of their host, relatively young, SNRs. In an evolutionary model [153], it was suggested that CCOs have high $B$ which was initially submerged. The re-emerging $B$ results in an increase of $\dot{P}$, and the CCOs will eventually evolve to join the RPP population. In an effort to reconcile the ages of pulsars and securely associated hosting SNRs, [57] showed that the apparently disparate classes of neutron stars, including the three pulsating CCOs, may be related to one another through the time evolution of the magnetic field [57].

**Magnetars**[2] are young neutron stars, typically characterized by ultra-strong $B$ well above the quantum critical field of $4.4 \times 10^{13}$ G. They are slow rotators with periods $P \gtrsim 2$ s, and often emit strongly in the X-ray band [see 186,229, for reviews]. Their quiescent luminosities often exceed their $\dot{E}_{SD}$, and thus it was theorized that the emission is powered by the decay of their enormous $B$ [354]. Beyond their significance for Galactic science, magnetars have been proposed to have connections with some of the most energetic and enigmatic phenomena in the distant universe, specifically Gamma-Ray Bursts (GRBs) and Fast Radio Bursts (FRBs). The latter, in particular, constitutes one of the most mysterious classes of cosmic objects. Recent propositions suggest a potential link between magnetars and certain FRBs, as illustrated by the case of the Galactic magnetar, SGR1935+2154. Swift's discovery of X-ray bursts from this active source was followed by the detection of Fast Radio Bursts, marking it as the first X-ray source associated with an FRB [39,61,266,402]. Consequently, zooming in on magnetars in our cosmic vicinity, particularly in X-rays, holds the promise of providing insights into the nature of these mysterious and distant FRBs.

Spectra of magnetars are well described by thermal BB from the hot surface and nonthermal radiation from the magnetosphere. The nonthermal emission is described by a soft power law at energies below $\sim 10$ keV but a dramatic spectral turn-over at energies above $\sim 10$ keV has been seen in many magnetars [e.g., 205]. The hallmark characteristics of magnetars are <1 s bursts and months-long outbursts during which their X-ray and/or soft gamma-ray emission increases by orders of magnitude [e.g., 187]. Because these emission properties (e.g., >10 keV spectral turn-over, outbursts, super-strong $B$, X-ray luminosity exceeding spin-down energy, overall lack of wind nebulae) are different from the other classes of neutron stars, it was thought that magnetars form a totally distinct class. However, magnetar-like outbursts from the typical high-$B$ (where $B$ just exceeds the quantum critical field) RPPs PSR J1846−0258 [107,206] and PSR J1119−6127 [400], discoveries of low-$B$ magnetars [e.g., 11,302,324,411], discoveries of wind nebulae around high-$B$ pulsars and magnetars [32,34,115,401] blurred the distinction significantly. These suggest that magnetars are neutron stars in a different stage of their magneto-thermal evolution [e.g., 295,373] or that the other classes of young neutron stars, like CCOs, will eventually show a magnetar-like behaviour (as for the CCO in the RCW 103 supernova remnant, [300]) or evolve into magnetars (e.g., [309]).

### 5.2. Important questions on Isolated Neutron Stars which AXIS can help address

Although various X-ray observatories such as *Chandra*, *XMM-Newton*, *Swift*, and *NuSTAR* have helped understand neutron stars in great detail during the previous two decades, a lot of important questions on neutron stars still remain unanswered. We list a few of them below.

---

[2]   https://www.physics.mcgill.ca/~pulsar/magnetar/main.html

1. Relation between the diverse classes of neutron stars: It was suggested that neutron stars born with different birth properties (e.g., $B$) evolve along different paths and exhibit diverse observational properties; e.g., it was suggested that magnetars evolve to become XDINSs [e.g., 172,185], and CCOs become RPPs [153]. The evolution effects may be observed as correlations between the emission and spin-inferred properties. Correlation studies have been performed with neutron stars in different classes [e.g., 13,154,237,281,414], which found correlations between $kT$ and $B_s$ and between $L_{BB}$ and $B_s$ in neutron-star samples taken from multiple classes, thermally emitting RPPs, XDINSs, and magnetars. These findings suggest that neutron stars have the same power sources [e.g., residual heat and decay of $B$; 281] but their relative contribution differs from class to class, providing important clues toward unification of neutron-star models. However, the sample sizes used in these studies were small, and the correlation significance was not very high. Moreover, relations between the properties (e.g., inferred from power-law fits) could not be measured precisely. These can be substantially improved by increasing the sample size of thermally-emitting neutron stars. *AXIS* with its large effective area and high angular resolution across the field of view (FoV) will discover more highly-absorbed (i.e., at large distances) low-$kT$ neutron stars (e.g., high-$B$ RPPs, XDINSs, CCOs, and transient magnetars in quiescence) that have not been detected by current X-ray observatories.

2. Long-term evolution of neutron stars over time scales of $>$ Myr: Magneto-thermal evolution models [275,295,373] have shown that long-term evolution of neutron stars depends on their physical properties such as internal $B$, core temperature, and the state of the dense matter. Note, however, that the samples used in these studies might be biased to high-$L_{BB}$ sources because low-$L_{BB}$ ones are difficult to discover. Nonetheless, in these models neutron stars in the diverse classes differ only in their birth properties and their evolution; these generate the observational diversities. The "neutron-star unification" schemes seem to successfully explain the $L_{BB}$ *vs* age trend [e.g., 295,373] and the magnetar-like outbursts from typical (high-$B$) RPPs [PSR J1846−0258 and PSR J1119−6127; 107,400] as due to magneto-thermal evolution effects [275,373]. The authors predicted that outbursts of older and low-$B$ (compared to magnetars) pulsars would be less frequent and have less energy. Then, two outbursts from the typical RPP PSR J1846−0258 on a time scale of ∼10 yr, which seems as frequent as ones in magnetars with much higher-$B$, are very challenging to explain. This may indicate the true $B$ of a neutron star is different from the spin-inferred $B_s$ [e.g., hidden toroidal component; 275], but only three magnetar-like outbursts were detected from two RPPs. So the unification models need to be scrutinized with larger samples of magnetar-like outbursts and low-$L_{BB}$ RPPs. Particularly useful for the study of the long-term evolution is a very young neutron star whose cooling can be detected on a time scale of order 10 years. Ho et al. [155] and Shternin et al. [331] suggested, using *Chandra* data spanning ∼19 yr, that the surface temperature of the ∼350-yr-old (inferred from the SNR expansion) CCO CXOU J232327.9+584842 in the Cas A SNR decreased at a 10-year rate of 2–3%. If so, the CCO provides a unique opportunity to probe initial cooling of newly-born neutron stars and fundamental physics, such as the presence of a neutron superfluid and proton superconductor in the star [265,331,332]. While the results of Ho et al. [155] were reproduced by Posselt & Pavlov [294], the latter noted that the changes of the fit-inferred temperature can alternatively be explained as due to systematic effects caused by, e.g., extra contamination of the ACIS filter beyond that accounted for in existing calibration information [280]. This controversy can be resolved by *AXIS* observations, which may also be able to measure potential temperature declines in other (slightly older) CCOs Ho et al. [see e.g. 155].

3. Short-term flux relaxation over time scales of years after magnetar outbursts: Magnetar outbursts can also tell us much about physics under extreme conditions; magnetar outbursts were suggested to be caused by crustal deformation and/or twist of the external $B$ [354]. Mechanisms of the outbursts and their relaxations are not yet well understood, but it is thought that the change of the thermal emission is related to crustal cooling [e.g., 201] whereas the change of the nonthermal emission is influenced by magnetospheric relaxation [30]. In the case of crustal cooling, the declining trends of $L_{BB}$ with time after outbursts can provide important insights into the dense matter in the crust [e.g., location of the energy deposition and core temperature; 12]. In particular, the late-time cooling trend (on time scales of thousands of days) is sensitive to the properties of the deep inner crust [e.g., nuclear-pasta phase; 161,283] and can help probe the state of the dense matter near the core [e.g., 81]. For such studies, observations of outburst relaxations of faint magnetars are desirable because for bright magnetars amplitudes of the relative flux enhancement and subsequent relaxation, significantly affected by neutrino emission [e.g., 282], are very small. As transient magnetars in quiescence may be very faint having a 0.5–10 keV flux of < $10^{-14}$ erg s$^{-1}$ cm$^{-2}$, current X-ray observatories have not followed the cooling trends down to very low-flux levels. *AXIS* will allow accurate characterizations of the late-time cooling trends of faint and transient sources and make it possible to probe properties of deep inner crusts of magnetars.

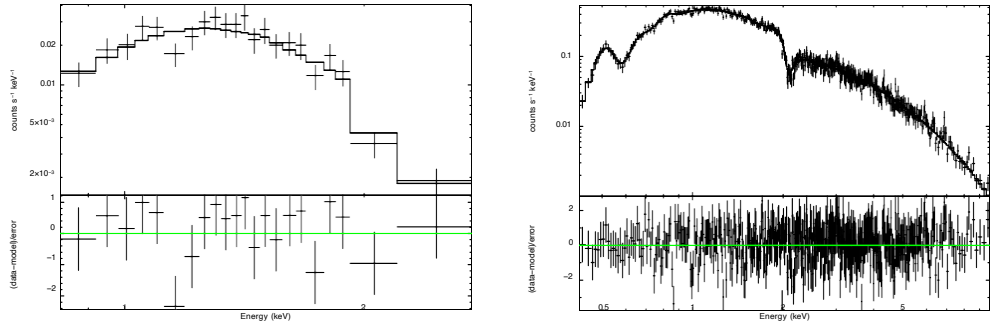

**Figure 11.** Spectra of the RPPs PSR J1734−3333 (left) and PSR J0205+6449 (right) simulated for a 20 ks and 50 ks *AXIS* observation, respectively. For PSR J1734−3333, we assumed a BB model having $N_H$ = $6.7 \times 10^{21}$ cm$^{-2}$, $kT$ = 0.3 keV and the unabsorbed 0.5–2 keV flux of $3.6 \times 10^{-14}$ erg s$^{-1}$ cm$^{-2}$ [260]. For J0205+6449, we based the 50-ks simulation on a BB+PL spectrum measured by 317 ks *Chandra* exposure [337].

### 5.3. AXIS *simulations for Isolated Neutron Stars*

As noted above (Section 5.2), *AXIS* can make significant contributions to our understanding of isolated neutron stars by characterizing the thermal emission of faint sources and by discovering more (and exotic) sources. Accurate characterizations of RPPs are crucial not only to population studies (e.g., correlation and cooling) but also to measurements of their PWN emissions since the pulsars' nonthermal emission can contaminate the PWN emissions in low-spatial-resolution observations. Some RPPs (including high-*B* RPPs) are observed to have only faint BB emission, and it was difficult to measure their $kT$ and $L_{BB}$ for population studies [e.g., 295,414]. The left panel of Figure 11 shows as an example a spectrum simulated for an *AXIS* observation of a faint high-*B* RPP [PSR J1734−3333 with the 0.5–2 keV flux of $3.6 \times 10^{-14}$ erg s$^{-1}$ cm$^{-2}$; 260] whose spectral parameters were not well constrained by a 125 ks *XMM-Newton* observation even with frozen $N_H$. A 10-ks *AXIS* observation can constrain the flux and $kT$ to within 6% and 10%, respectively (for frozen $N_H$), and 20-ks data can put tight constraints on the $N_H$ (20% uncertainty) as well as the spectral parameters (30% and 10% uncertainties for the flux and $kT$, respectively). For some RPPs, their BB emissions are not well constrained due to contamination by the bright nonthermal emission of the pulsar and PWN. Moreover, pile-up effects in *Chandra* observations and large PWN+SNR background in *XMM-Newton* (and *Chandra* CC-mode) data have prohibited precise characterizations of the BB component. These can be overcome by *AXIS* thanks to its large effective area, and high temporal and angular resolution. BB emission has been certainly detected in some young and bright RPPs [e.g., PSR J0205+6449; 337], suggesting that there are more such sources in which the BB emission is undetected due to the strong nonthermal contamination. As 'young' RPPs with small $L_{BB}$ can provide important clues to the thermal evolution of neutron stars [e.g., 295,373], discovering more young and low-$L_{BB}$ RPPs is crucial to constraining the evolution models (Section 5.2). From *AXIS* simulations for PSR J0205+6449, we found that its faint thermal emission, which is swamped by a nonthermal power-law component (pulsar + PWN), can be detected with $3\sigma$ confidence by a 10 ks *AXIS* observation, and $kT$ and $L_{BB}$ can be measured to within 5% and 30% with a 50 ks observation (Figure 11). This verifies that *AXIS* can characterize faint BB emission of RPPs and will help scrutinize the magneto-thermal evolution models.

*AXIS* can also increase the sample size for correlation studies. XDINSs and magnetars can be identified by *AXIS* as the 3–17 s pulsations can be easily detected. With a time resolution of less than 50 ms, *AXIS* could also measure thermal and non-thermal pulsations from RPPs with spin periods of several hundred ms; for subarray observations, even faster spinning RPPs could be detected, such as the 65 ms PSR J0205+6449. The spatial resolution of *AXIS* will be important to reduce non-pulsed emission from any nearby PWN and/or SNR, and the large effective area of *AXIS* will allow discoveries of distant and highly-absorbed sources. Figure 12 shows results of simulations for 10 ks *AXIS* observations of a source having $kT$ = 0.1 keV and $L_{BB} = 10^{33}$ erg s$^{-1}$ (for XDINSs, CCOs, and transient magnetars in quiescence). The pulsed fraction and distance to the source were varied, and we assumed that the absorbing column density $N_H$ increases by $10^{22}$ cm$^{-2}$ per kpc. The simulations show that a 10-ks *AXIS* observation will be able to detect the pulsations of a neutron star at $\leq 3$ kpc if its pulsed fraction is greater than 20%. Two of the currently known XDINSs (<500 pc) have a pulsed fraction of $\approx 20\%$ [178], and extrapolating this to the larger space volume we anticipate *AXIS* will discover $\sim 70$ similar XDINSs. Although this estimation is very rough, it is almost certain that the high angular resolution over the large FoV of *AXIS* will allow serendipitous discoveries of many XDINSs and quiescent magnetars. The latter likely having higher $kT$ and nonthermal emission may be more easily detected.

The 100–400 ms pulsations of CCOs might be detectable by *AXIS*. Identification of a newly-discovered source as a CCO can be assisted by measurements of the emission spectrum [$kT$=0.15–0.47 keV; 9] and the location (SNR association). We investigated how well a 10-ks *AXIS* observation would 'discover' a BB-emitting source within an SNR. The middle panel of Figure 12 shows detection significance for BB emission with $L_{BB} = 10^{33}$ erg s$^{-1}$ for various values of $kT$ and distance. The results suggest that *AXIS* can make a $10\sigma$ detection of a CCO at $\geq 10$ kpc if its $kT$ is greater than 0.3 keV. The detection criterion ($10\sigma$ instead of $3\sigma$) we used here mitigates the influence of possible background by the PWN and SNR. The high detection sensitivity of *AXIS* will also allow meaningful follow-up studies of magnetar relaxation after an outburst down to a very low-flux level; e.g., a 10-ks *AXIS* observation could have detected the low-$B$ magnetar SGR 0418+5729 at 2 kpc, had its flux decayed to a lower level (absorbed 0.5–10 keV flux of $6 \times 10^{-15}$ erg s$^{-1}$ cm$^{-2}$; $10\sigma$ detection) than the historical minimum of $2 \times 10^{-14}$ erg s$^{-1}$ cm$^{-2}$ [302].

*AXIS* will be able to resolve the uncertainty of the cooling rate of the CCO CXOU J232327.9+584842 in the Cas A SNR, whose observation requires the spatial resolution of *AXIS* due to the diffuse X-ray bright emission around the CCO. In the right panel of Figure 12, we show current measurements of the surface temperature of the CCO and simulation results for three 50 ks *AXIS* observations. For the simulations, we assumed that the temperature drop is astrophysical in origin [155,331] and is not caused by systematic effects [e.g., 294]. A 50-ks *AXIS* observation will constrain the surface temperature to within $\sim 0.1\%$, and three such observations with a cadence of $\leq 2$ yr are sufficient to tell whether or not the surface temperature of the source actually decreases with time. This could have significant implications on long-term magneto-thermal evolution models and constraints on fundamental physics. We note that the *Chandra* and *AXIS* correlated measurements are subject to systematic uncertainties of the different instruments. However, even if we make a single *AXIS* temperature measurement and find it significantly lower than the average value of all the *Chandra* data, or lower than simply the early *Chandra* values where the contaminant build-up is not significant, then this could be used to argue for cooling. Furthermore, if we make several *AXIS* measurements over a few years, as shown in Figure 12, and find using only these *AXIS* datapoints that there is a temperature decline comparable to that derived from *Chandra*, this would further suggest that the coolong is real. Finally, we note that *AXIS* will potentially measure cooling in other young neutron stars over its mission lifetime.

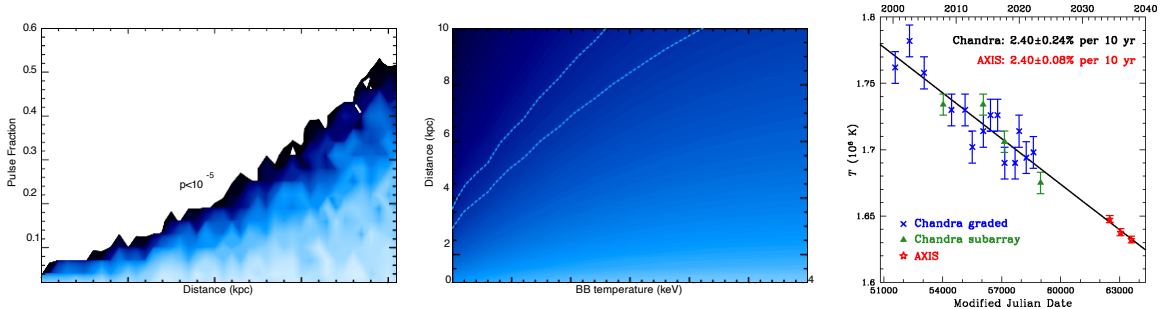

**Figure 12.** Simulations of *AXIS* observations for detections of pulsations (left) and faint sources (middle), and for measurements of the surface temperature of the CCO in the Cas A SNR (right). *Left and middle*: Simulations for 10-ks *AXIS* observations were performed for a thermally emitting (BB) source having $kT = 0.1$ keV (varying in the middle panel) and $L_{BB} = 10^{33}$ erg s$^{-1}$ with $N_H$ increasing by $10^{22}$ cm$^{-2}$ per kpc. The absorbed 0.5–10 keV flux is $2.5 \times 10^{-14}$ erg s$^{-1}$ cm$^{-2}$ (for $kT = 0.1$ keV at 1 kpc) and decreases inversely proportional to distance squared. The white region in the left panel is where the pulsations could be detected with the chance probability less than $10^{-5}$, and the contours in the middle panel denote detection significances for reference ($5\sigma$ and $10\sigma$). *Right*: We extrapolated from previous measurements of the spectra of the Cas A CCO reported by Shternin et al. [331], simulated three 50-ks *AXIS* spectra, and plotted the temperature inferred from the *AXIS* data (red points).

Thanks to its large effective area, *AXIS* will also allow detections and studies on spectral features from magnetars. To date, several spectral features from magnetars have been detected during quiescence as well as during bursts (see e.g. [14,106,167,301,347, line detections at various energies during bursts of several magnetars] [356, variable absorption lines throughout the rotation phase of the magnetar SGR 0418+5729]). These features are often interpreted as proton cyclotron features, as a result of photons emitted from the hotspot scattering at the cyclotron frequency in the magnetosphere. However, the extreme magnetic and gravitational fields in the region are expected to broaden such line features, making them harder to detect. Larger effective area instruments with sufficient timing and energy resolution, therefore, are crucial to be able to detect stronger spectral signatures of magnetars. Magnetar burst line detections promise insight into the exact causes of bursts and emission mechanisms. On the other hand, during quiescence, varying spectral lines in phase-resolved magnetar spectra are thought to be linked to the geometry of the magnetic field lines surrounding the emitting region [356]. More recent work shows that hotspot emission and corresponding magnetic field geometry in quiescence can be determined by constraining the exact shape of the spectral line (i.e., line width and depth) concurrently with the line energy using phase-resolved spectroscopy [194]. Overall, spectral lines also yield an estimate of the magnetar's magnetic and gravitational field as well as the size of the emitting region.

We performed simulations to check for the capabilities of *AXIS* in detecting spectral features of magnetars during quiescence and bursts. For quiescent emission, we focused on phase-resolved spectroscopy to test how well *AXIS* could detect varying spectral features throughout the rotational phase of the magnetar compared to other current/upcoming telescopes. For all simulations, we created fake spectra using the *fakeit* function on XSPEC v12.13.0 [19] with the latest public effective area and response of the respective detector. For persistent emission, we set all models and corresponding parameters to the line detections between of SGR 0418+5729 reported in [356]. For bursts, we set model parameters to the 1998 burst precursor line detection of SGR 1900+14 in [347], but scaled continuum normalization to obtain a count rate of ∼ 35k in 0.2 seconds duration to represent a burst-like scenario. We then fit the simulated spectra for various detectors, fixing the quiescent emission continuum model to the reported phase-averaged best-fits in [356] and SGR 1900+14 Hydrogen Column Density ($N_H$) to the best-fit value in [347]. We freed all remaining parameters. Note that for now we did not include pile-up effects in these simulations.

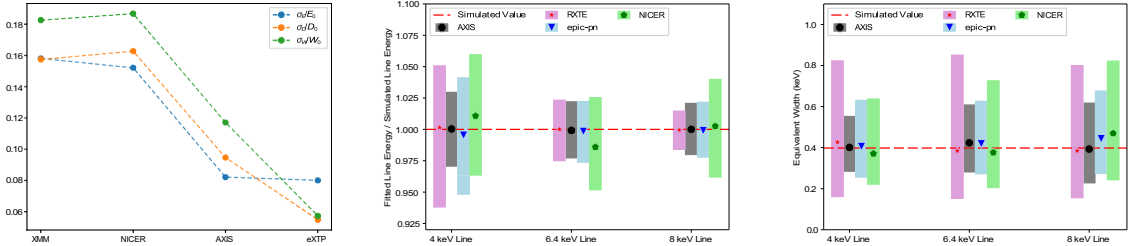

**Figure 13.** *Left Panel:* Results of quiescent magnetar phase-resolved spectroscopy simulations as average percent errors (i.e., 1σ errors divided by the fitted parameter value) for line energy (in blue), line depth (in orange) and line width (in green) for various telescopes. Simulations were conducted based on the models and parameters in [356]. *Middle and Right Panels:* Average fitted per simulated line energy (*Middle)* and fitted line width (*Right)* for three line energies simulated based on the original 6.4 keV burst spectral line detection of SGR 1900+14 in [347]. Error bars reported are 1σ parameter errors averaged over 100 iterations. Red dotted lines indicate simulated values.

In Figure 13, we present the results of our simulations for magnetar spectroscopy. On the left panel, we plot the average percent errors (i.e., 1σ errors divided by the fitted parameter value) for line energy (E), line depth (D) and line width (W) for quiescent emission. Note that all values reported here are averaged over simulations for ten spectral lines detected between the phases 0−0.3 for SGR 0418+5729 in [356]. We see that, compared to current detectors (i.e., *XMM-Newton* EPIC-pn and *NICER*), *AXIS* can detect all spectral line parameters with over a two-fold reduction in percent errors, and performs similarly to the upcoming eXTP mission. This shows that *AXIS* will increase our capabilities in detecting magnetar spectral lines in quiescence. Moreover, we will be able to probe hotspot emission and corresponding magnetic field geometry thanks to the capabilities of *AXIS* in constraining all spectral line parameters concurrently (see [194] for a more detailed description on how constraining these line features would enhance our understanding of magnetar field line geometry).

In the middle and right panels of Figure 13 we also plot the results of our simulations for magnetar bursts. Here, values reported for each detector are fitted values with corresponding 1σ upper and lower error bars averaged over 100 iterations. To test for a hypothetical spectral burst line at higher and lower energies, we re-run the simulation for a 4 keV and 8 keV line in addition to the original 6.4 keV line detection in [347], while keeping all other parameters the same. We find that overall *AXIS* provides the most accurate as well as precise measurements of all line parameters during the burst compared to current instruments. Since magnetar bursts tend to be extremely energetic, the large effective area of *AXIS* would also be useful to detect spectral lines during bursts (as opposed to several previous detections only during burst precursors likely due to pile-up limitations). Therefore, we conclude that *AXIS* could provide a useful tool in understanding more about magnetar bursts through spectral feature detections.

*5.4. Summary*

Isolated neutron stars have been intensely studied during the previous decades by X-ray observatories, which have significantly improved our understanding of neutron stars. However, there are still a lot of things that are not well understood. As we demonstrated above, *AXIS* will discover more neutron stars and characterize their emission properties accurately, enabling meaningful population studies from which a unification of the neutron-star classes may be achieved. Although not detailed above, *AXIS* studies of individual neutron stars will also help in understanding physics under extreme conditions; e.g., accurate measurements of the spectra of CCOs and XDINSs can reveal the nature of the absorption features (cyclotron *vs* atomic transition) commonly seen in CCOs and XDINSs. With AXIS, such features may be found from more isolated neutron stars in different classes [e.g., like one in SGR 0418+5729; 356]. As demonstrated here, *AXIS* will deepen our understanding of isolated neutron stars in many ways.

# 6. Compact Objects: Accretion-Powered Compact Objects

*6.1. X-ray binaries*

The *AXIS* Galactic plane survey will represent a leap forward in understanding the population of X-ray binary systems in our Galaxy.

### 6.1.1. Low Mass X-ray Binaries (LMXBs)

Jonker et al. [174] estimated that within the footprint of the *Chandra* Bulge survey, which covered 12 square degrees, that there was a population of approximately 530 LMXBs (the vast majority of which were in quiescence). Assuming a similar source density throughout the *AXIS* GPS footprint, the survey should observe a field containing 2000–3000 LMXBs. As noted in Jonker et al. [174], the mean X-ray luminosity of quiescent LMXBs (qLMXBs) is around $10^{33}$ erg/s, and their survey depth was optimized such as to be sensitive to about half of these sources. With a depth of $3 \times 10^{-15}$ erg/s, the *AXIS* GPS would completely detect the hypothetical population of LMXBs presented in the *Chandra* Bulge survey [174], including those below its detection threshold. Additionally, the bright half of sources would be detected with high significance, allowing for high-quality spectral characterization, as well as time domain studies of the X-ray variability in these sources. Because the *AXIS* GPS will be a cadenced survey conducted with on the order of ten epochs spaced out over 5 years, we also expect to be sensitive to X-ray transients, which may allow for the detection of outbursting LMXBs which elude all sky X-ray monitoring facilities.

### 6.1.2. Ultra-Compact X-ray Binaries (UCXBs)

In addition to being a powerful facility for identifying ultracompact accreting white dwarf binaries, we expect the *AXIS* Galactic plane survey footprint to host 2000–3000 UCXBs with BH/NS accretors based on the densities presented in Jonker et al. [174]. Notably, these systems have an order of magnitude lower luminosity in their quiescent states than longer-period UCXBs, making the sensitivity of the *AXIS* GPS a particular asset in identifying these systems. Many such systems may undergo eclipses in the optical (as the disk is occulted by the donor), and thus cross matches of the *AXIS* GPS point source catalog with surveys such as Roman and LSST could be a powerful way to identify these rare systems. Additionally, some of these systems, like their white dwarf binary counterparts, will be luminous in millihertz gravitational waves, and will be detected by LISA. Thus, *AXIS* could be used to provide rapid localization of the GW source by significantly cutting down on the number of possible electromagnetic counterparts (since most of these sources will be undergoing stable mass transfer, and thus are likely to exhibit a positive $\dot{P}$ in LISA, indicating that one should look in the X-ray to find the counterpart). Additionally, the *AXIS* GPS and deep surveys of globular clusters may directly reveal more rare UCXBs via periodicity searches in the X-rays, such as the candidate 28 minute orbital period BH+WD UCXB, 47 Tuc X9 [25].

*6.2. Ultra-Luminous X-ray Sources (ULXs)*

A large number of XRBs with luminosities that exceed $10^{39}$ erg/s have been discovered in nearby galaxies. These are commonly referred to as Ultra-Luminous X-ray sources [ULXs, see recent review; 193]. It has been speculated that ULXs offer evidence for the existence of intermediate-mass BHs (IMBHs, with $M \sim 10^2 - 10^4 \, M_\odot$; [70]). Alternatively, ULXs can be products of super-Eddington accretion onto a stellar-mass compact object [328]. The latter scenario has been favored by the presence of photoionized nebulae around ULXs that may be naturally explained by strong outflows combined with the large intrinsic X-ray luminosity [296,341,343]. More specifically, theoretical models for ULXs invoke super-critical accretion discs, where advection and outflows play a key role in shaping the disc structure. However, the smoking gun for super-Eddington accretion was the discovery of pulsations from M82 X-2, a system with luminosity of $10^2 \, L_{\mathrm{Edd}}$, demonstrating that stable accretion onto NSs at super-Eddington rates is possible [21]. This discovery introduced a new category of systems, the so-called Ultra-Luminous X-ray Pulsars (ULXPs). This realization has fueled a search that led to the discovery and study of more ULXPs in recent years [e.g., 53,103,170,308,323]. Furthermore, based on spectral similarities between non-pulsating and pulsating ULXs, there is now compelling evidence that a significant fraction of ULXs may actually host strongly magnetized NSs [$B > 10^{12}$ G; 198,378]. Based on surveys and efforts of multiple observatories more than 2000 ULX candidates are now known [e.g. 202,377]. However, apart from the ULXPs the nature of most of the other sources remains elusive. *AXIS* can help push forward the study of ULXs. Apart from the obvious increase in identified ULXs that will result from the increased effective area and reduced PSF size compared to current observatories (i.e., *XMM-Newton*, *Chandra*, *Swift*), *AXIS* will greatly increase our understanding of the known ULXs in both spectral and timing properties.

In terms of spectral studies of ULXs *AXIS* can deliver up to 4 times more counts than *XMM-Newton*, while due to more flexible pointing capabilities monitoring surveys of nearby galaxies (like M51) could enable study of spectral variability of ULXs during super-orbital and/or outburst cycles [e.g. 131,210,211].

ULXs also exhibit long-term variability that sometimes can be quasi-periodic and related to some precession of the system [e.g. 77,200,210], while other systems might show sudden drops in flux that may be related to propeller transition and thus be evidence of magnetized neutron stars [232]. *AXIS* observations during low flux phases of ULXPs would enable tests of the mechanism behind these transitions in more systems.

Meanwhile, the timing resolution of *AXIS* (i.e., smaller than 0.2 s) would be enough to search for pulsations from many more ULXs, or track the spin evolution of ULXPs. We note that most ULXPs have spin periods close to 1-3 s, with very low pulsed fractions (10-20%). The exception is NGC 300 ULX-1 where its spin period evolved between 126 s to 16 s within a couple of years, while its pulsed fraction was more than 50% [368]. To quantify the capabilities of *AXIS* for such timing studies, we performed simulations on the ULXP source M51 ULX-7, a 2.8s pulsar on a 2-day binary system located in the outskirts of the spiral galaxy M51a at a distance of 8.6 Mpc [308]. Moreover, the flux of the system varies between $1\text{-}10\times10^{39}$ erg/s, showing a super-orbital modulation with a 40 d period and some evidence of propeller transitions [369]. Based on archival *XMM-Newton* observations, pulsations are not always visible and pulsed fraction is on the order of 10-15%. With AXIS, we could detect about 3.5 times more counts in a single visit compared to *XMM-Newton*, while the background contamination (due to diffuse emission and other point sources in M51) would be minimal compared to that in an *XMM-Newton* observation (20% or even more in counts). By performing simulations we found that assuming a pulsed fraction of 0.2, pulsations would be detected with as little as 5000-7000 counts which may be collected in a 10-20 ks *AXIS* observation throughout the super-orbital modulation. The major advantage of *AXIS* would be that the spin period can be detected in a fraction of the orbital period, thus pulsations would not be smeared by the Doppler effect. In contrast, the search for pulsations with *XMM-Newton* would require exposures over 100 ks, and would require acceleration searches [308].

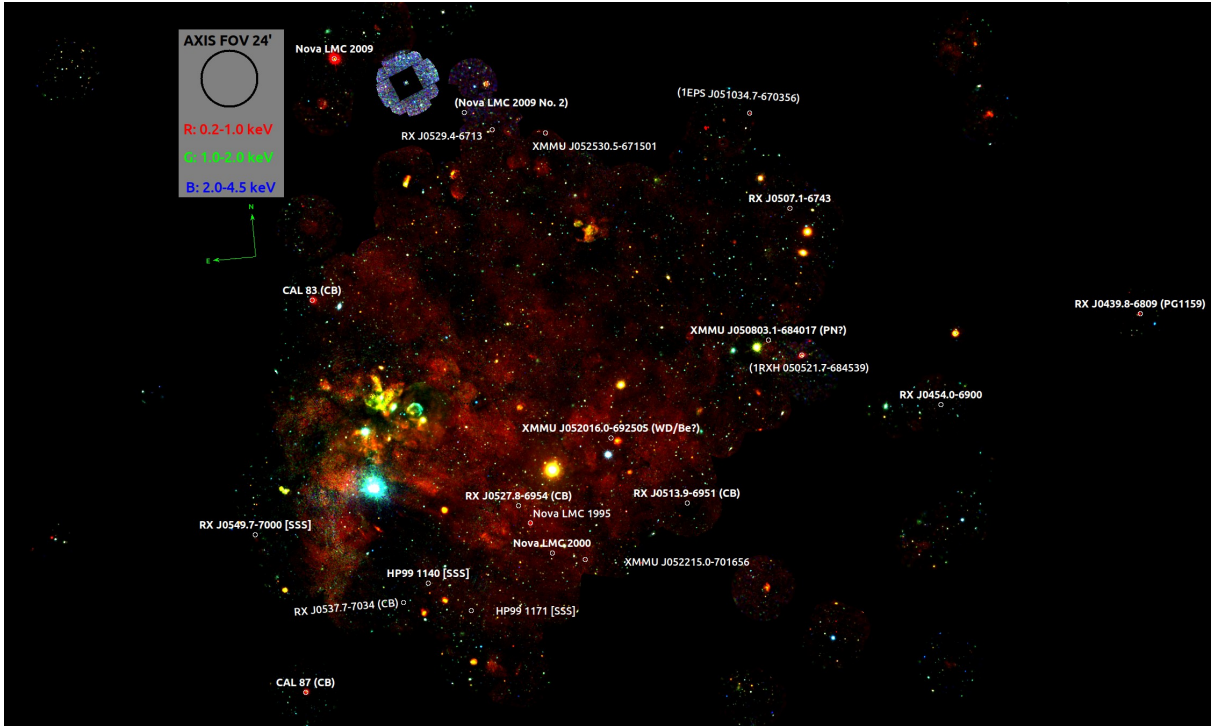

**Figure 14.** X-ray mosaic image (Red:0.2-1.0 keV, Green:1.0-2.0 keV, Blue:2.0-4.5 keV) of the LMC based on *XMM-Newton* observations (credit: *XMM-Newton* Large survey of LMC, PI F. Haberl). Regions mark the known close-binary supersoft sources in the LMC, while diffuse soft X-ray emission (i.e., <1 keV) is evident throughout the galaxy. The *AXIS* field of view (24′ diameter) is also marked in the top left corner.

### 6.3. Super Soft Active Galactic Nuclei (SS AGN)

Super Soft Active Galactic Nuclei (SS AGN) are soft X-ray excess dominated AGN [223,333]. They are identified by very steep X-ray spectra (photon index $\Gamma > 3$) and high X-ray luminosities ($L_X > 10^{41}$ erg s$^{-1}$) [349]. They host higher-mass intermediate to lower-mass supermassive black holes with $M_{BH} \sim 10^4 - 10^6 \, M_\odot$ [311] and are likely to be the missing link between the ULXs and normal AGN. Sacchi et al. [311] identified and confirmed only five such AGN from the *XMM-Newton* catalog of serendipitous sources [4XMM-DR9, 384]. The *AXIS* deep sky surveys will significantly increase the current sample size of SS AGN.

SS AGN emit X-ray photons predominantly below 2 keV. It is thus unknown whether they intrinsically lack the hard X-ray emission or the current X-ray telescopes are not sensitive enough to detect the weak hard X-ray emission from the corona. For a given radius, the inner disk temperature of SS AGN is higher compared to typical AGN disks, which could potentially impact the coronal geometry and/or emission properties of SS AGN. Thus the high sensitivity of *AXIS* will be crucial in determining the true nature of hard X-ray emission and corona in SS AGN.

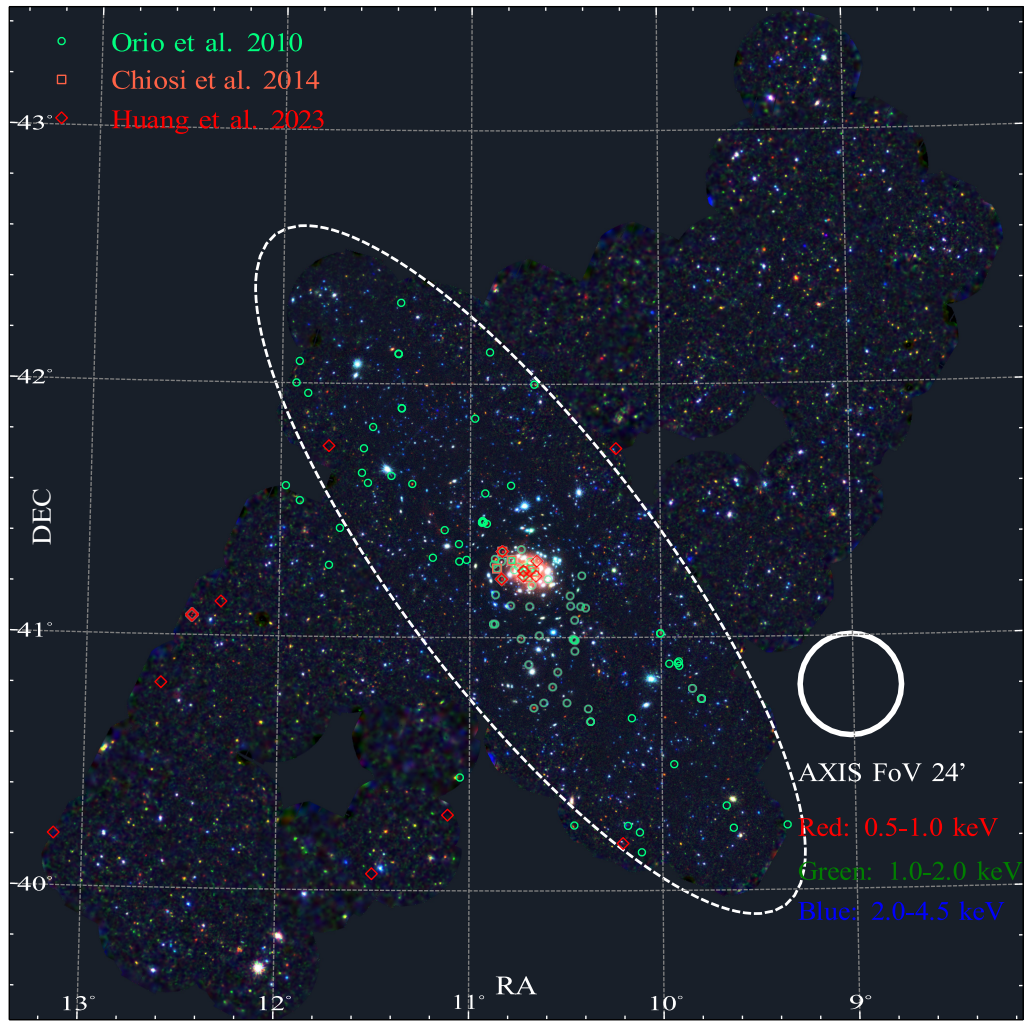

**Figure 15.** X-ray mosaic image (Red:0.5-1.0 keV, Green:1.0-2.0 keV, Blue:2.0-4.5 keV) of M31's disk and halo based on *XMM-Newton* observations (credit: the "An XMM-Newton*XMM-Newton* View of the ANdromeda Galaxy as Explored in a Legacy Survey (New-ANGELS)" project, PI J. Li; [162]). The dashed ellipse marks the optical $D_{25}$ region. Different small regions mark the identified SSSs in different works [62,162,263]. *AXIS* field of view (24′ diameter) is also marked in the lower right corner.

*6.4. Super Soft Sources (SSSs)*

First discovered in the Magellanic Clouds by the Einstein Observatory (HEAO-2)'s early exploration of the X-ray sky (see figure 14 for the currently known LMC population), the class of objects we now call supersoft X-ray sources (SSSs) are classified by their very soft, blackbody-like spectra (with $kT \sim 20–100$ eV) and luminosities ranging from $10^{35}–10^{38}$ erg/s [125,126]. Based on this energy release and their inferred radii, these objects were identified as accreting white dwarfs undergoing nuclear-burning near their surface [177,366]. It was soon found that within a narrow range of accretion rates ($\approx 1–4 \times 10^{-7} M_\odot/yr$, consistent with thermal timescale mass transfer) steady, persistent nuclear burning of all accreted material was possible. Above this range, optically-thick winds may drive an outflow [134] and on longer timescales, re-inflate the white dwarf envelope to giant dimensions [55]. For accretion rates below the steady-burning threshold, matter accumulates in the partially degenerate envelope until triggering a thermonuclear eruption, giving rise to classical and recurrent novae [e.g., 59,298]. Following the nova outburst, an accreting white dwarf may sustain a post-nova supersoft phase whose duration depends on the mass of the WD and the rate at which it's accreting [340].

It is possible that supersoft X-ray sources are, instead, powered by super-Eddington dynamical timescale/unstable mass transfer onto a stellar-mass black hole or neutron star companion, prior to common-envelope event. In such scenarios, the powerful outflows accompanying the rapid mass transfer will inflate a compact, radio-synchrotron-bright "hypernebula" [341]. The dense disk outflows do not allow the disk photons to emerge directly from the disk surface, instead, they emerge after multiple scatterings from the fast wind/jet photosphere at much larger radii. This reduces the effective temperature of the disk emission to $10 \sim 100$ eV thus enabling hypernebulae to be candidates for ultraluminous supersoft X-ray sources. These sources, if jetted, could be sources of Fast Radio Bursts if the jet is pointed along our line of sight [342]. Furthermore, the protons accelerated at the jet termination shock of hypernebulae could interact with the beamed soft X-ray disk photons and produce high-energy neutrinos. Integrated over volume and time, this could potentially make SSSs one of the significant contributors to the background extragalactic high-energy neutrino flux as seen by IceCube [343].

SSSs have received intensive scrutiny due to their possible role as progenitors of Type Ia Supernovae (see above), as well as the role novae [108] and rapidly-accreting white dwarfs [82] may play in the origin of the elements. Accreting white dwarfs also provide a vital benchmark in our understanding of the stability and evolution of mass transfer in binary stars, and in particular, the possible mechanisms underlying the mysterious common envelope phase, a vital step in the formation of virtually all compact binaries and gravitational wave sources. This closely links the understanding of SSSs with the goals prioritized by the Astro2020's Report of the Panel on Stars, the Sun, and Stellar Populations, in particular:

**G-Q2: "How does multiplicity affect the way a star lives and dies?"**

Making progress on understanding the still-little-understood formation, evolution, and ultimate fate of SSSs will require extraordinary **sensitivity in the soft X-ray band**, superior **angular and timing resolution**, and **dedicated monitoring** campaigns working in synergy with surveys in other wavebands. This is an approach identified as particularly essential by Astro2020's New Messengers and New Physics priority, noting "*the power of near-continuous monitoring in the X-ray, gamma-ray, optical, infrared, and radio bands has been dramatically demonstrated over the past two decades.*"

*AXIS* will provide an absolutely essential tool in unraveling these mysteries, driving forward our understanding of accreting white dwarfs across several orders of magnitude in accretion rates, as well as their behaviour on timescales from seconds to decades. Its relatively wide field of view and soft X-ray sensitivity will allow future guest observer programs on *AXIS* to quickly and efficiently map the SSSs of nearby stellar populations. Assuming a typical column density $N_H = 10^{21} cm^{-2}$, and that a reliable spectral measurement of the effective temperature of a given source would require ∼200 counts, we find that *AXIS* will be able to detect any steady-burning SSS with a white dwarf mass above ∼ 0.8$M_\odot$ in M31 within a 20ks exposure (see the known SSSs around M31 in Fig. 15), and any such white dwarf above ∼ 0.55$M_\odot$ in the LMC in a 1 ks exposure (see Figure 16). At these depths, *AXIS* would be able to carry out a comprehensive survey of M31 in ∼18 pointings, and cover a broad swath of the central region of the LMC in ∼60 pointings, providing a powerfully complete census of both persistent and post-novae SSSs; repeating all or part of this campaign on an approximately annual basis would in turn provide an invaluable history of their long-term behaviour, resolving key questions such as the mechanism(s) underlying post-nova mass loss [390]; the true numbers of such objects and their possible role in some Type Ia supernova explosions [340]; and their role in the origin of *i*-process and other elements [82]. At the same time, monitoring short term pulsations observed in persistent and post-nova SSSs (on timescales of tens to hundreds of seconds) can disambiguate whether these pulsations are driven by g-mode oscillations in the outer envelope [253,390] or are instead evidence of a rotating hot spot on the white dwarf surface [see, e.g., 253,370, for further discussion], in either case potentially providing a novel means of inferring the mass and possible growth of the accreting white dwarf, and the physics underlying its continued accretion. Looking further afield, *AXIS* observations of nearby spiral galaxies would provide an invaluable probe of SSS evolution while complementing a host of ancillary Galaxies science, from understanding stellar and black hole feedback to the physics of the circumgalactic medium. In particular, ∼200 ks total exposures following-up the *Chandra* Survey of Nearby Galaxies (11 galaxies at distances of ∼4–13Mpc) [192] supplemented by, e.g., additional objects from the PHANGS survey [207] could provide a comprehensive census of the most massive accreting nuclear-burning white dwarfs (i.e., those closest to the Chandraskehar mass) in a range of environments spanning different star formation histories and metallicities. These results are vital to our emerging picture of the evolution of Type Ia supernova progenitors over cosmic time, and a critical test of our understanding of binary stellar evolution in the local Universe.

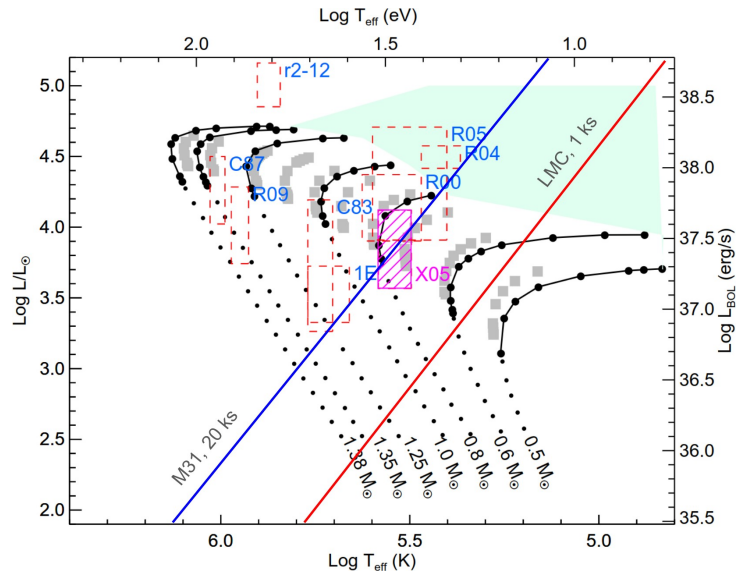

**Figure 16.** Detection limits for a given blackbody luminosity and temperature, assuming $N_H = 10^{21} cm^{-2}$ and two assumptions about distance and total integration time: 20ks for an object in M31 (blue line) and 1ks for an object in the LMC (red line). Several SSSs are marked in the plot with their observed properties RXJ0925.7-4758 (R09), CAL 83 (C83), 1E 0035.4-7230 (1E), RX J0019.8+2156 (R00), RX J0439.8-6809 (R04), RX J0513.9-6951 (R05), CAL 87 (C87) and 1RXS J050526.3-684628 (X05). Also shown are the stable-burning models of [389] (gray squares), and the stable-burning (connected large black dots) and nova (small black dot) white dwarf models of [258]. (Original figure credit [370])

Fundamental to the discovery potential of *AXIS* in uncovering the evolution of accreting compact object populations will be its strong synergies with other great observatories of the next decade, as well as ongoing efforts today. With its relatively wide field of view and high angular resolution, *AXIS* will be uniquely well-matched to the Cosmological Advanced Survey Telescope for Optical and uv Research (CASTOR), a proposed 1m UV-optical space telescope with 0.15" resolution and ~0.25 square degree field of view that will reach AB~27 in ~600s, as well as providing grism and multi-object spectroscopy, planned for launch by the end of the decade [73]. Together, CASTOR and *AXIS* would provide an ideal combination for complementary surveys in UV & X-rays of nearby binary populations, especially in crowded fields; enable multi-wavelength fitting of close binary supersoft source spectra; and allow monitoring of massive companions of HMXBs & accretion disks. With both missions anticipating dedicated surveys of the Galactic plane, and likely the LMC and M31 as well, CASTOR and *AXIS* would provide a uniquely powerful pairing. By leveraging the results of presently ongoing deep, multi-epoch spectroscopic surveys, in particular the Sloan Digital Sky Survey (SDSS)-V [199], *AXIS* could further probe the role that accreting compact objects may play in contributing to the ionizing background in galaxies [391,392].

# 7. Pulsar Wind Nebulae (PWNe)

## 7.1. PWNe as most extreme particle accelerators and nearby laboratories for fundamental physics

Pulsar-wind nebulae (PWNe) harbor some of the most extreme particle accelerators known to exist in the Galaxy – young rotation-powered pulsars powering ultrarelativistic magnetized winds (see [104,179, 307,312,313]). Studies of synchrotron emission, dominating PWNe spectra from radio to MeV gamma-rays, provide a window into the inner workings of these remarkable relativistic magnetized plasma laboratories. In particular, high-resolution X-ray images and spatially resolved spectroscopy can elucidate the following aspects:

- collisionless shock physics in relativistic magnetized outflows (e.g., by resolving termination and bow-shock morphologies and particle SED evolution);
- kinetic particle escape, conditions in ISM and its magnetic field, and propagation of ultrarelativistic particles in ISM, including the positron distribution in the Galaxy;
- magnetic reconnection and turbulence in magnetized plasma;
- particle acceleration mechanisms operating in relativistic magnetized outflows and pulsar electrodynamics (e.g., pair cascade physics and spin-dipole axis alignment);
- supernova explosion physics leading to neutron star kicks and misalignment between the pulsar spin and magnetic axis;
- massive star evolution by establishing the progenitors of rotation-powered pulsars through establishing connections between pulsars and properties of their host SNRs.

## 7.2. High-Resolution imaging and spectral index mapping of PWNe

The morphologies and spectra of PWNe [182] depend on the anisotropy of the wind, its magnetization, particle acceleration efficiency, magnetic field strength, and pulsar velocity. These intertwined dependencies can be disentangled by (1) resolving the PWN structure, (2) connecting PWN properties with the pulsar properties and the ambient medium properties (including the host SNR), and (3) increasing the number of PWNe detected in X-rays to enable meaningful population studies.

While many pulsars have reliably measured spin-down energy loss rates ($\dot{E}$), it is much more difficult to determine some other important parameters potentially influencing the morphologies, radiative efficiencies, and spectra of PWNe. Two of these, the angle $\zeta$ between the pulsar's spin-axis and the observer's line of sight and the inclination angle $\alpha$ between the spin and magnetic dipole axes (if the field is indeed largely dipolar). However, these angles are notoriously difficult to measure, but can be constrained by comparing theoretical models of pulsar magnetospheric emission with the *Fermi* LAT GeV light curves and radio light curves (e.g., [57,276,383]). However, different magnetospheric models can give differing predictions. Since pulsar winds are intrinsically anisotropic, many PWNe exhibit prominent equatorial and polar components (e.g., torii and jets) which reflect symmetry with respect to the pulsar spin-axis (see, e.g., the Crab and Vela PWNe in Figure 17). Resolving these PWN features often allows one to identify the symmetry axis (rotation axis) and, hence, $\zeta$ (see [254,255]). Independent determination of $\zeta$ breaks the degeneracy between the models and helps to determine the one which is consistent with the data. Some PWNe display nearly-axial symmetry, but identifying the equatorial and polar outflows can be challenging (e.g., MSH 11–62 and G21.5–0.9 shown in Figure 19). The anisotropy of pulsar winds can lead to anisotropic PWN spectra, flow speeds, magnetic fields, and cooling trends, requiring high-resolution imaging, spatially-resolved spectroscopy, and spatially-dependent models accounting for wind anisotropy. It is plausible that the lower values of $\alpha$ would result in a highly magnetized wind with a stronger polar outflow (compared to the equatorial torus) and also perhaps a reduced particle acceleration efficiency and lower PWN luminosities (see modeling by [45]). However, a larger sample of PWNe with spatially-resolved morphologies and spectra is needed to perform informative population studies and test theoretical predictions for models of relativistic magnetized outflows from pulsars.

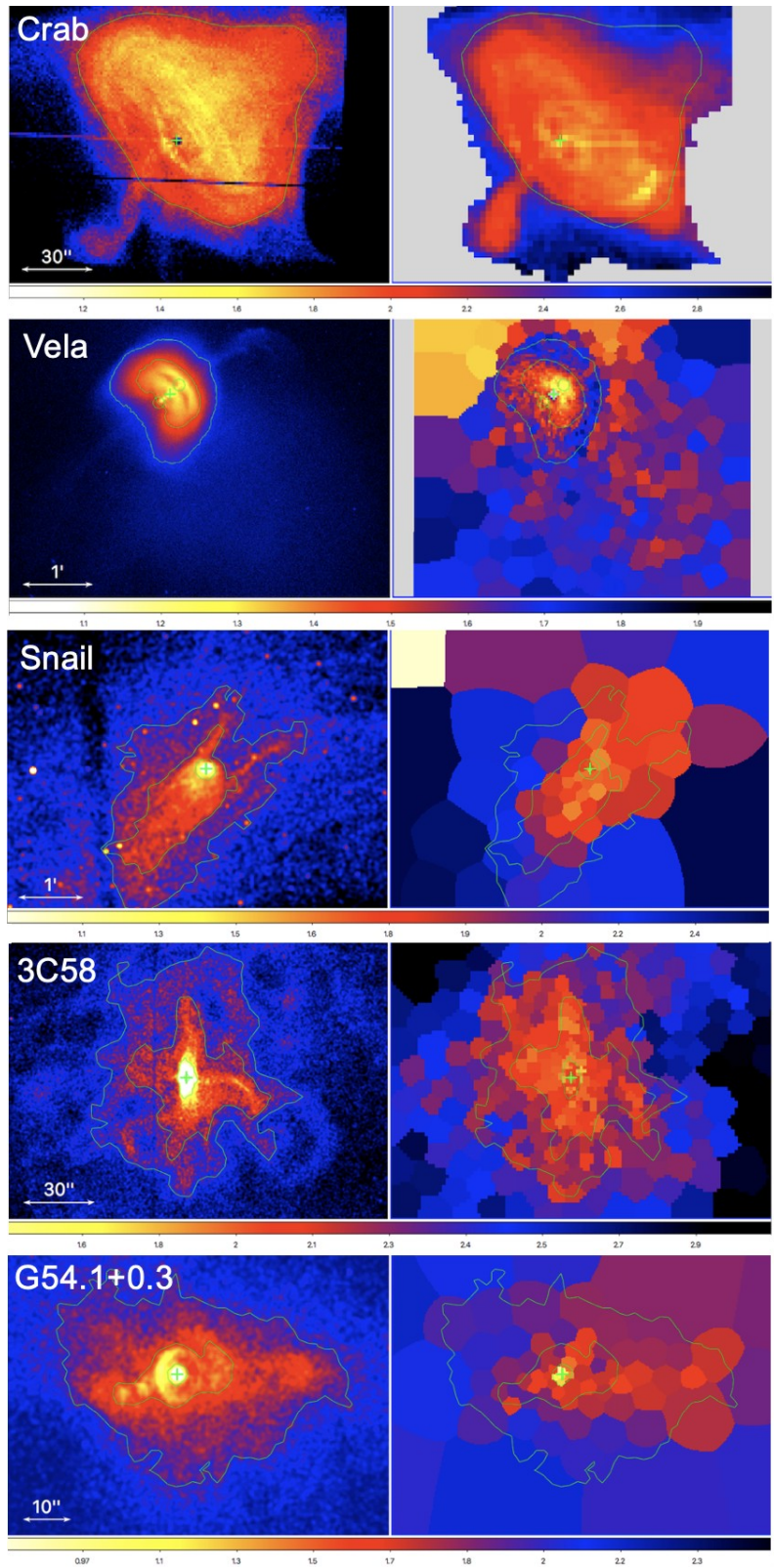

**Figure 17.** *Left panels: Chandra* ACIS images (in 0.5–8 keV) of selected, well-resolved PWNe. *Right panels:* Adaptively-binned spatially-resolved spectral maps for the PWNe shown on the left. The color bars represent the photon index Γ measured in the 0.5–8 keV band. The green contours are shown for illustrative purposes, and the green crosses mark the pulsar positions. The gray-colored areas in panels 1 and 2 were excluded from mapping. The Crab spectral map is adopted from [238]. Adopted from [180].

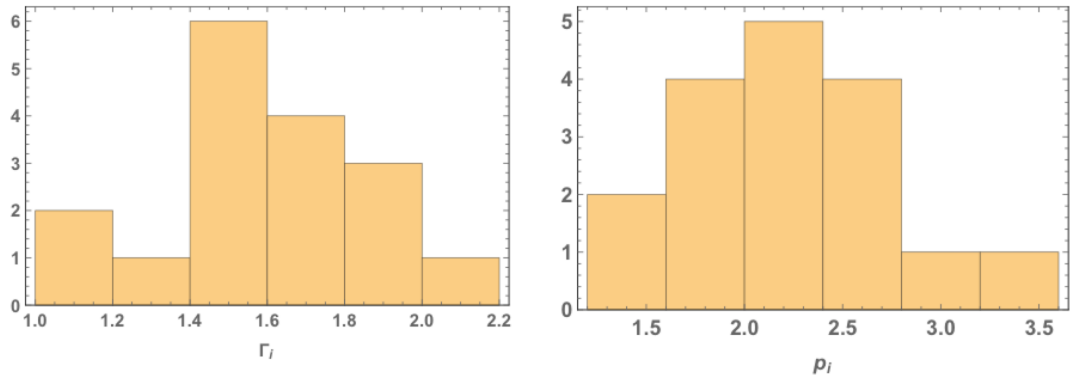

**Figure 18.** Histograms of the photon and particle indices, $\Gamma_i$ and $p_i = 2\Gamma_i - 1$, for a sample of *Chandra*-studied PWNe. The bin width corresponds to the average measurement uncertainty. Adopted from [180] and using their Table 2 sample.

It is worth noting that a handful of Chandra-discovered PWNe surrounding highly energetic ($\dot{E}_{sp} \geq 5 \times 10^{36}\,\text{erg s}^{-1}$) pulsars are extremely compact and faint in X-rays, at odds with PWNe being calorimeters of active rotation-powered pulsars (e.g., [20,117,316]). *AXIS*, thanks to its unprecedented sensitivity and high-angular resolution, will enable us to probe this poorly understood class and find more such, previously missed, objects which will help address the zoo of PWNe and the impact of the pulsar wind magnetization on the morphology and spectral properties of these objects.

Lastly, *Chandra*'s imaging resolution has been crucial to interpret the X-ray polarization maps recently obtained with the Imaging X-ray Polarimetry Explorer (IXPE). For example, IXPE's observations of the Vela PWN [395] showed that the nebula's magnetic field is highly ordered, and symmetrical around the direction of the pulsar's motion, *following the arcs imaged with* Chandra (see Fig. 2 in [315]). Such an 'order' in Vela challenges models predicting that unstable flows or turbulence play a key role in accelerating particles in ultra-relativistic winds. In the 2030's, *AXIS* will be synergistic with X-ray polarization missions aiming at studying the particle acceleration process in relativistic plasma.

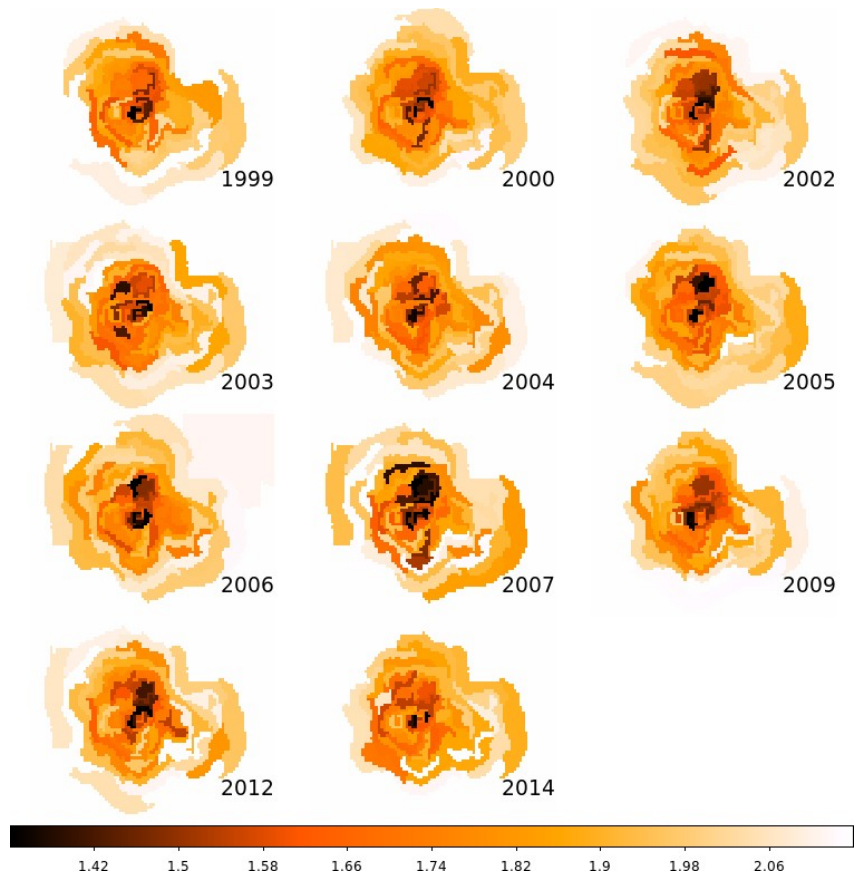

**Figure 19.** *Chandra* spectral map of the PWN G21.5–0.9 between 1999 and 2014 showing variability of the photon index on small scales. These minute changes in spectral index became apparent even in cases where brightness images alone do not suggest any underlying changes, revealing the spectral map analysis technique as a new way to view variability of PWN winds on sub-parsec scales. Figure credit:[128].

### 7.2.1. PWNe Variability

While *Chandra* revealed jets and torii structures in PWNe, with spatially dependent spectral indices, it has also revealed variability of PWN structures on small sub-arcseconds scales in the brightest PWNe (e.g., [196,272]). Observing variability within PWNe has been traditionally limited to cases of significant brightening, or the few instances where transient features are interpreted in terms of intrinsic motion or associated with variability from the pulsar. Jet and torus morphology are also only visible in cases of differing brightness with respect to the surrounding nebula and favourable alignment with our line of sight. A recent study of a small sample of young and bright PWNe, observed during the lifetime of the *Chandra* mission, found evidence of previous unknown variability using the spectral map analysis technique, opening a new window into viewing PWNe [128].

Such studies have been limited to the youngest, brightest and heavily observed-by-*Chandra* objects. AXIS, thanks to its unprecedented sensitivity and imaging resolution over a large field of view, combined with its low background, will not only increase the significance of detection for the already studied objects, but will also expand the sample to the fainter and older population of PWNe, including bow-shock nebulae, while taking advantage of observations already carried out by *Chandra*. Modelling and comparisons with 3D MHD simulations are also needed to understand the origins of this newly revealed form of variability.

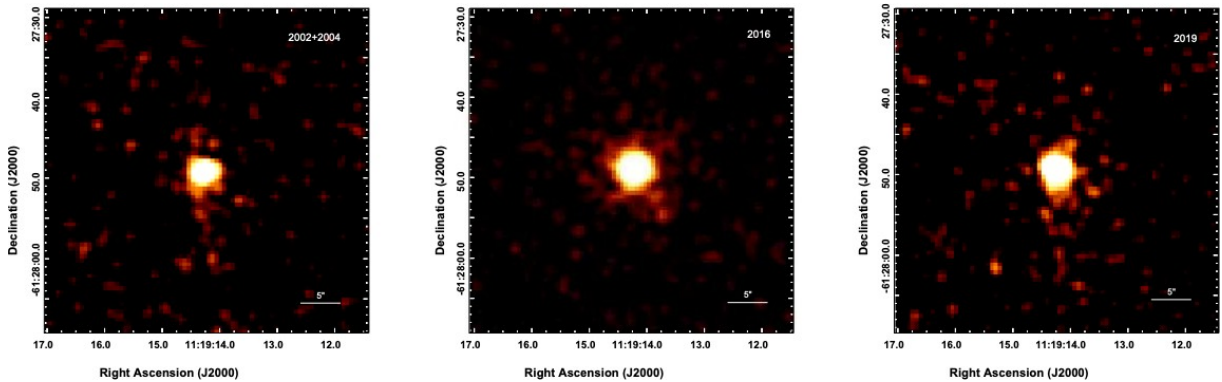

**Figure 20.** *Chandra* images of the compact ($\sim 3'' \times 6''$) pulsar wind nebula, surrounding the high-*B* 'schizophrenic' pulsar PSR J1119–6127. The images show the nebula when the pulsar was in quiescence (left), during a magnetar-like outburst detected in 2016 (middle), and after it returned to quiescence (right). The PWN changed both in its morphology and spectral properties showing the impct of the magnetar outburst on the pulsar's environment. This special case illustrates the power of high-resolution imaging in detecting such rare cases of a pulsar/magnetar wind nebula in an object that is thought to be a rotation-powered pulsar but occasionally behaves like a magnetar. Figure credit:[33].

In particular, and as an example, for G21.5–0.9 – the canonical example of a Crab-like PWN, although with a shell-like component found after accumulating some 500 ks of *Chandra* time [227] and used as a calibration target for *Chandra* – spectral variability on the angular scale of a few arcseconds and timescale of 1 year or less was found [128]. This is illustrated in Figure 19. *AXIS* will provide the necessary statistics to confirm this variability, and search for variability in other PWNe. A characteristic spectrum from a $3'' \times 3''$ square of the G21.5–0.9 PWN requires more than 130 ks of observing time to yield the $\sim 5,000$ photon counts required to constrain the observed variability to better than $5\sigma$. *AXIS* will require a mere 20 ks. The brief exposures combined with the large field of view of *AXIS* will allow measurements of spectral variability in the central PWNe of young composite-type SNRs.

### 7.3. Wind nebulae around high-B pulsars and magnetars

The investigation into whether highly magnetized neutron stars, magnetars and High-*B* pulsars (HBPs) with magnetic fields (*B*) exceeding $4 \times 10^{13}$ G, can power PWNe remains an intriguing and open question. The formation of PWNe typically requires a relativistic particle outflow within a strongly magnetized and confining medium. While magnetars have been proposed to generate steady or post-outburst particle outflows [140], identifying PWNe around magnetars in particular poses a challenge due to their relatively high X-ray luminosity and significant interstellar absorption, leading to the formation of a dust scattering halo around them. Typically, the X-ray luminosity of PWNe associated with Rotation-Powered Pulsars (RPPs) constitutes only a small fraction of their spin-down energy. In the case of magnetars, the spin-down energy is usually much smaller than that of RPPs, with typical values around $10^{33}$ erg s$^1$. Assuming a similarly small X-ray to spin-down luminosity ratio for steady PWNe around magnetars and HBPs, deep high-resolution X-ray observations become crucial to identify any associated PWN. This underscores the potential of the *AXIS* mission for such investigations.

Currently, only a few wind nebulae have been observed around these objects (see e.g., [1,312]), and the question persists regarding whether they are powered by rotation, magnetism, or a combination of both (e.g., [1,34,120]). Notably, *Chandra* facilitated the discovery of a compact PWN around the High-B pulsar, PSR J1119-6127, known to be a RPP until it exhibited a magnetar-like behavior in 2016 [33,319] (see Fig. 20). *Chandra* also played a crucial role in identifying a compact and faint wind nebula candidate around the youngest known radio-loud magnetar Swift J1818.0-1607 [32]. While several other compact magnetar wind nebula candidates exist, the most compelling evidence for a magnetar wind nebula surrounds Swift J1834.9–0846 [401]. AXIS is poised to expand this sample, efficiently map the environment around these objects, and disentangle the dust scattering component from the PWN component. Such studies not only aids in constraining their properties but also facilitates the search for any associated Supernova Remnants.

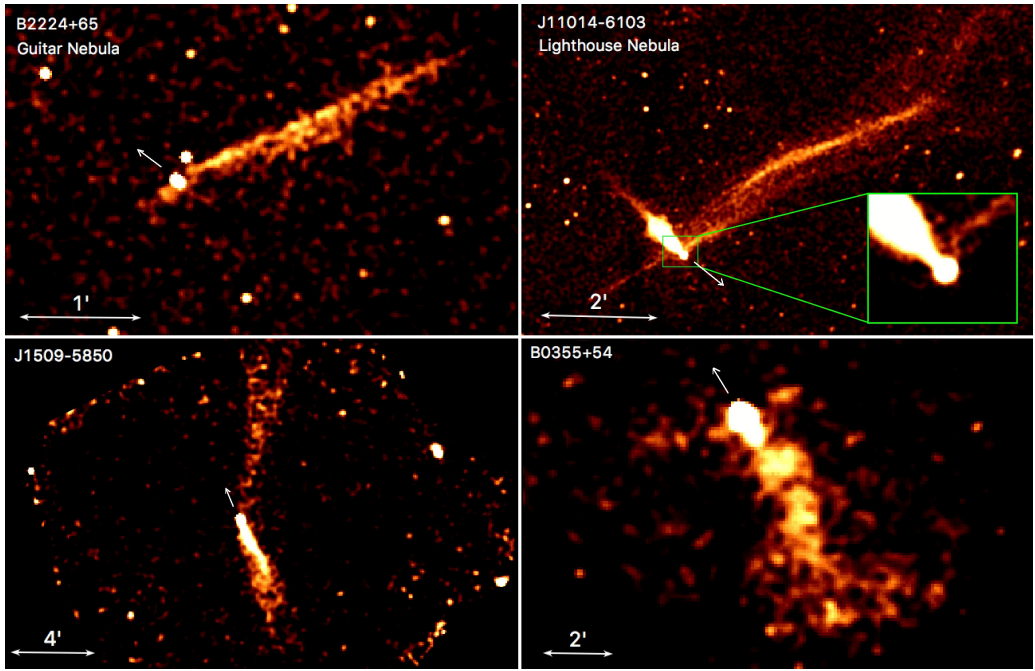

**Figure 21.** *Chandra* ACIS images of supersonic PWNe displaying tails and misaligned outflows. The white arrows show the directions of pulsar's proper motion. Adopted from [184].

### 7.4. Natal kicks and supersonic pulsar winds

The most telling sign of the acceleration mechanism is the slope (shape) of the spectral energy distribution (SED) of accelerated particles and its dependence on other (e.g., pulsar or ambient medium) parameters. However, pulsar wind particles suffer radiative energy losses causing the evolution of the SED with distance from the pulsar. Therefore, to determine the particle SED injected at the termination shock and probe the mechanism for pulsar wind propagation into the SNR or ISM, one must obtain statistically significant, spatially-resolved PWN spectra and compare them to models (see Figure 17, 18 and [128,129]).

The "natal kicks" that pulsars receive in supernova explosions lead to high pulsar velocities with the average $v_{psr} \sim 400$ km s$^{-1}$ [156]. This implies that most pulsars remain inside their host SNRs only for a few tens of kyrs and then escape into the interstellar medium (ISM) where the speed of sound $c_{ISM} \ll v_{psr}$ (typically, $c_{ISM} \sim 3 - 30$ km s$^{-1}$, depending on the ISM phase). Therefore, in the ISM the pulsar motion becomes supersonic (Mach number $\mathsf{M} \equiv v_{psr}/c_{ISM} > 1$), and the ram pressure from the oncoming ambient medium strongly modifies the PWN appearance leading to the formation of extended pulsar tails (see [184,307] for reviews on supersonic PWNe). Although in supersonic PWNe the structures formed by the anisotropic wind (e.g., jets and tori) are deformed by the ram pressure, in some cases, the torus-jet structure can still be identified in high-resolution images (see, e.g., J1509–5850, Geminga, B1706–44, and B0355+54 in Figure 2 of [184]).

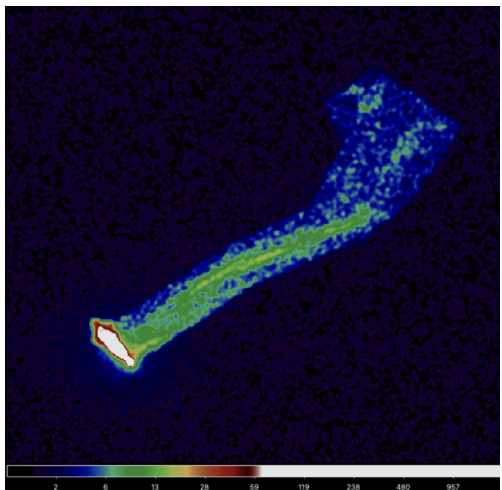

**Figure 22.** *AXIS* simulation (300 ks) created using `sixte` (with the *Chandra* image as input). The scalebar is in units of counts arcsec$^{-1}$. With the same exposure time, *AXIS* will collect $\approx$ 8 times more counts than *Chandra*. Note, that with 1″ PSF FWHM *AXIS* can recover all structures indicated by *Chandra* observations within the extended outflow. It will also resolve the morphology of the compact PWN (tail and the head where the extended outflow originates as two faint streams).

In recent deep X-ray observations, a new type of structure has unexpectedly been discovered in some supersonic PWNe. Extended, elongated features, *strongly misaligned with the pulsar's direction of motion*, are seen originating from the vicinity of four pulsars (see Figure 21 for examples). The misaligned orientation of these features is surprising because for a fast-moving pulsar, one would expect all of the pulsar wind to be confined within the tail (which these PWNe exhibit as well). A hypothesis where relativistic particles populating the misaligned features escape from the compressed apex of the bowshock, where magnetic field reconnection can help the escape, has been put forward by [27] and further developed by [29] and [261]. It is important to stress that in this scenario the misaligned outflows are an entirely kinetic phenomenon, which makes them very different from the pressure-confined jets of tails. Since only the most energetic particles are expected to escape from the bowshock apex one test of this hypothesis would be the absence (or weakness) of the cooling (spectral softening) along the extended misaligned features. This kind of signature can be efficiently probed with *AXIS* for the brightest structures (e.g., the Lighthouse nebula; see Figure 22). The observed appearances of the misaligned outflows should reflect the ambient ISM structure illuminated by synchrotron emission from the leaked pulsar wind particles. Thanks to its sensitivity and large field of view, *AXIS* is expected to discover more of these remarkable structures around fast-moving pulsars elucidating ultra-relativistic particle (including positrons) escape and transport in the ISM.

### 7.5. TeV Halos and Galactic PeVatrons

A related topic is the Galactic PeVatron candidates associated with evolved PWNe inside composite type SNRs (e.g., [46,49,292]), and extended (with scales of tens of pc) 'TeV halos' (also dubbed relic PWNe) discovered recently in high-energy gamma-rays by Cherenkov imaging telescopes around some middle-aged pulsars [6,217,325]. Due to their large extent (in TeV) and confusion with many X-ray sources in the field, these TeV sources often include alternative possible sources of relativistic particles (such as SNRs and XRBs) but the leading hypothesis is that the TeV emitting particles are pulsar wind particles which mostly have cooled too much to detect their synchrotron radiation in X-rays, but the up-scattered (by the IC mechanism) background radiation is still seen in TeV, owing to a longer cooling time for this process. Sensitive observations with *AXIS* may pick up the faint synchrotron emission for PWNe associated with the pulsars powering TeV halos (see Fig. 23 as an example) and reveal the expected fading with the distance from pulsar due to the radiative cooling. Such observations will be highly synergistic with LHAASO's current hunt for PeVatrons [50], radio and TeV observations from upcoming SKA and CTA by allowing to map out the evolution of the entire multiwavelength SED of the relic PWNe and enabling direct comparison with the particle transport model predictions.

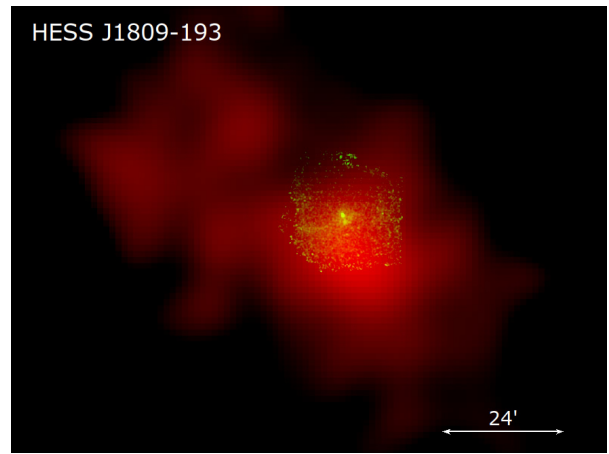

**Figure 23.** *Chandra* and H.E.S.S. image of the HESS J1809−193 source, showing a TeV halo surrounding a pulsar and its compact X-ray nebula. The green color is the *Chandra* image in the 0.5–8 keV band showing the PWN associated with PSR J1809−1917 and the red color is high-energy gamma-rays (> 0.27 TeV) from HESS J1809−193. Figure credit:[132,197].

*AXIS*'s excellent angular resolution and large effective area, combined with the low detector background, make it possible to resolve (spatially and spectrally) the complex anisotropic structures of PWNe, such as bow shocks, wisps, knots, rings, jets, arcs and combinations thereof (see Figure 17). On larger scales (∼0.1–10 pc), *AXIS* can resolve structures such as long diffuse tails directed opposite to the pulsar's motion, and strongly misaligned (with the pulsar's direction of motion) extended outflows (see, e.g., Figure 22 and Figure 9 in [184]).

In summary, the PWNe study strategy for *AXIS* can be three-fold:

- Deep observations of a relatively small sample of bright and large PWNe (most of which have been already studied with *Chandra*) with the goal of obtaining high-quality images and spatially-resolved spectra.
- Targeted deep searches for wind nebulae around HBPs, particularly in the aftermath of any magnetar-like outbursts, and magnetars themselves, serving as a strategic initiative to investigate the environment of these objects. Such efforts are poised to establish connections with various classes of neutron stars and are likely to yield insights into the elusive Fast Radio Bursts (FRBs).
- Searches of extended emission around young and middle-aged pulsars, as well as searches for non-thermal extended sources in the serendipitous Galactic plane observations to increase the overall PWN population. The number of PWNe detected in X-rays is of the order of 100 as of Sep. 2023 (ref. SNRcat [96]). We expect that *AXIS* can double this sample by discovering new PWNe in X-rays and by following up on PWNe discovered in radio (e.g., with SKA, ngVLA, and their prototypes) as well as in TeV by CTA. The actual number will, of course, depend on the length of the mission operation.

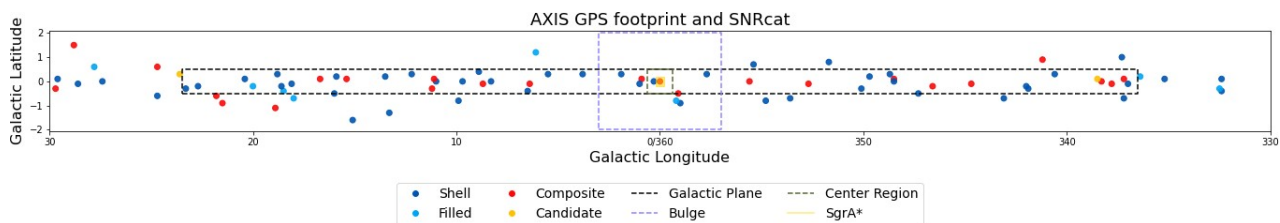

**Figure 24.** The known Supernova Remnants and Pulsar Wind Nebulae (from the High-Energy Catalogue of Supernova Remnants, SNRcat) covered within the *AXIS* GPS footprints. Shells are shell-like remnants, Filled-Centres are PWNe or plerions, Composites have both a shell-like and a centrally-filled component and include both plerionic (pulsar-powered) and thermal composites (also known as Mixed-Morphology SNRs). Candidate SNRs and PWNe are also included. As of July 2023, the total number of SNRs in the Galactic Plane Survey coverage is 50: 1 Candidate, 17 Composites, 2 Filled-Centre and 30 Shell-type. The total number of SNRs in the bulge = 9: 3 Composites, 1 Filled-centre and 5 Shell-type. The total number of SNRs in the Galactic Centre is 2: 1 Composite and 1 Shell-type.

## 8. Supernova Remnants

Supernova Remnants (SNRs) are among the most fascinating astrophysical objects in the Universe. They result from energetic explosions marking the death of stars and the formation of strong shocks driving hot, X-ray emitting, low-density plasma associated with both shocked circumstellar/interstellar gas and stellar ejecta. SNRs impact the chemical enrichment and evolution of galaxies, accelerate cosmic rays to extremely high energies, and those resulting from core-collapse explosions make the most magnetic, compact, and dense objects in the Universe: neutron stars (NSs) and black holes. As such, SNRs represent the best nearby laboratories to study extreme astrophysics that cannot be achieved even in the most extreme particle accelerators on Earth, as well as relativistic outflows and jets that are ubiquitous in Astrophysics. These objects have not only driven scientific breakthroughs, technology development and interdisciplinary connections, but they also fascinate the public and young people.

To date we know of some 400 SNRs in our Galaxy, many of which have been imaged and studied in detail in the X-ray band, thanks to modern observations with *Chandra* and *XMM-Newton*. See SNRcat [96][3] for the high-energy, regularly updated, public database of SNRs, PWNe and associated compact objects; and Fig. 24 for the known SNRs and PWNe covered in the proposed Galactic Plane Survey (see §2) footprints. We note that the number of currently known SNRs and PWNe is a factor of a few to 10 times less than the expected number of Galactic SNRs (and depends on the SNR lifetime and Galactic SN rate). Sensitive surveys are contributing to adding new candidates, but *AXIS* unique capabilities will help identify and discover faint SNR structures, the youngest SNRs and compact PWNe and magnetar wind nebulae – thus opening a new window to address the 'missing SNR problem'.

The past decade has in particular helped address many questions related to SNR formation and evolution; however our studies of SNRs have been limited to the brightest or closest objects to us, and we have so far touched only the tip of the iceberg. *AXIS*, thanks to its unprecedented sensitivity in the 0.5–10 keV band, excellent PSF over a large FoV, and low background will enable us to dramatically increase the

---

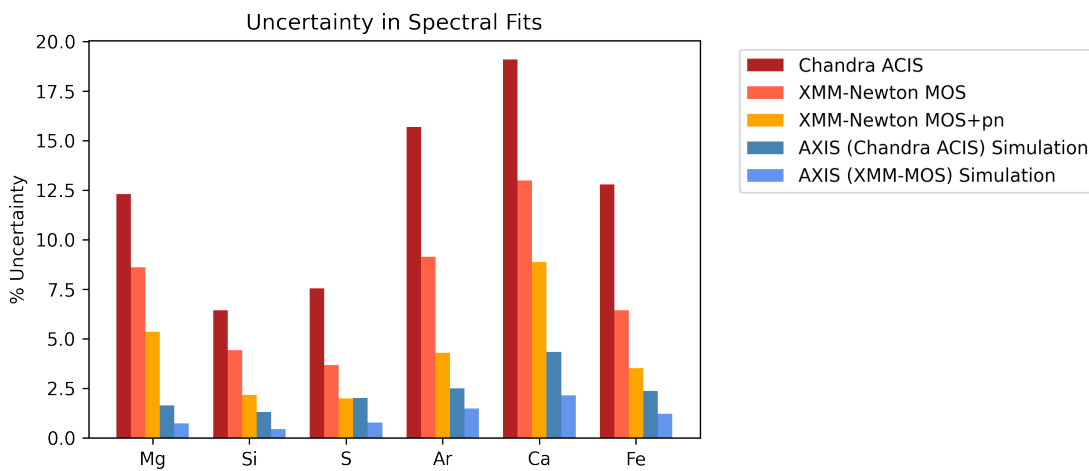

**Figure 25.** Abundances and associated uncertainties for a controversial SNR (in terms of its type) inferred from fitting X-ray spectra obtained with *Chandra*, *XMM-Newton* and simulated *AXIS* data with similar exposure times. This is shown as a representative example of the power of *AXIS* to better constrain the metal abundances in comparison to *Chandra* and *XMM-Newton*.

population of faint SNRs in our galaxy and nearby Universe, and through targeted observations, tackle some of the most fundamental outstanding questions in the field that are relevant to Astro2020. Some of these pressing questions are highlighted below.

A fundamental outstanding question about the evolution of massive stars is: **What are the progenitors and explosion mechanisms of supernovae? Which stars make neutron stars and which ones make black holes?** Among the core-collapse explosions that make neutron stars, there is a diversity that leaves behind a zoo of engines as described in §5. For Type Ia SNe, commonly used as standard candles for cosmology, there is an ongoing debate about their explosion mechanism: single degenerate versus double degenerate (see 2.1); and for some type Ia double-degenerate models, the lower mass white dwarf can survive the explosion leaving imprints in the supernova remnant for thousands of years after the explosion [97]. Supernova remnants are ideal targets to probe the physics and aftermath of SN explosions: they provide us with a nearby laboratory to directly probe the compact engines and the supernova ejecta which can glow in the X-ray band for tens of thousands of years post the supernova explosion.

## 8.1. Mapping SNRs to their SN progenitors

SNRs play a pivotal role as primary sources of products resulting from stellar and explosive nucleosynthesis. Nucleosynthesis, a complex field involving various processes operating in diverse environments and phases of stellar evolution, lacks comprehensive observational tests, particularly for the most energetic processes involved. Among these, the mechanisms responsible for the production and ejection of iron (Fe) and other elements in the Fe-group during the explosions of both core-collapse and thermonuclear supernovae are among the least scrutinized components of the nucleosynthesis framework. Young SNRs, especially those with an age of less than 10,000 years, are particularly crucial for delving into the chemical elements generated through a multitude of debated processes. X-ray studies, in particular, provide critical tools for testing the nucleosynthesis theories, specifically concerning the production of Fe-group and intermediate-mass elements in both core-collapse and thermonuclear supernovae. These studies not only provide vital empirical tests for nucleosynthesis models but also offer essential insights into the explosion processes intrinsic to supernovae, given that the production of Fe-group elements lies at the core of these cataclysmic events (see e.g., [22,102,165,166,374]).

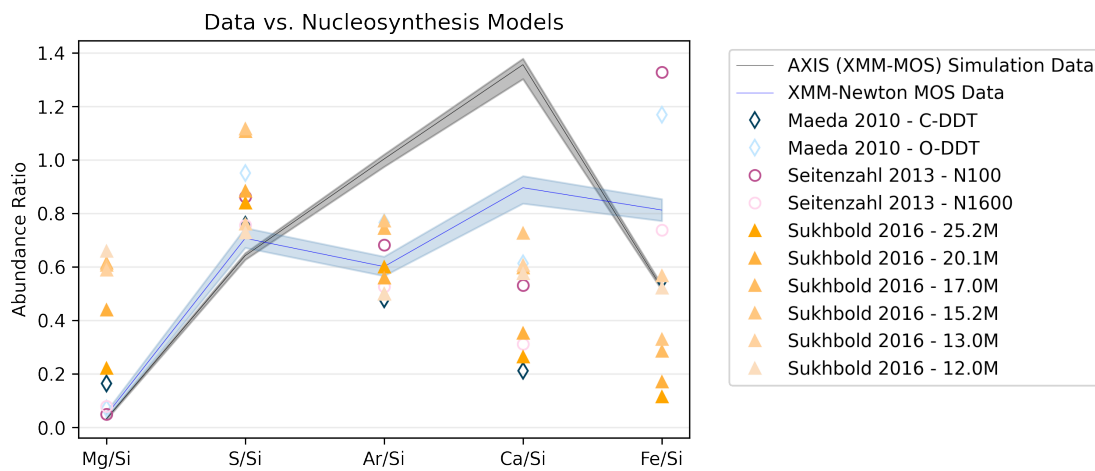

**Figure 26.** *XMM-Newton* and simulated *AXIS* data versus nucleosynthesis models (Ia and core-collapse) for a SNR of controversial type. This is shown as a representative example of the power of *AXIS* to better constrain abundances while revealing degeneracies in spectral fits and comparison to a suite of nucleosynthesis models.

While typing SNe has been traditionally done through their optical light curves, X-ray observations of SNRs (in our Galaxy and nearby galaxies) provide a powerful tool to infer their SN type, probe their energetics and environment, and map them to their progenitors. In the context of nucleosynthesis, core-collapse remnants are widely believed to be the primary contributors to the production of oxygen (O), neon (Ne), magnesium (Mg), silicon (Si), and sulfur (S). While they generate a diverse array of elemental species, including those in the Fe-group, the core-collapse process is particularly influential in shaping the final amounts of these alpha-elements. The quantities of these elements are directly proportional to the initial mass of the progenitor star and depend on the specific nucleosynthesis model employed.

X-ray astronomers, in collaboration with theorists and numericists, have in the past few decades used (young or ejecta-dominated) SNRs to infer the SN type or the physical mechanism within a SN type, through morphological studies (e.g, [97,98,214,216,264]), iron K-alpha line centroid mapping (e.g, [269,397]), and spatially resolved spectroscopic studies compared with nucleosynthesis or dynamical radiative evolutionary models aimed at measuring the metal yields and SN explosion properties (e.g., [118,352], more below); all methods require high-resolution imaging in X-rays. For the latter effort, spatially resolved X-ray spectra are first fitted with thermal plasma models, then the abundances of elements (from O through Fe and Ni) are inferred and compared to those predicted from a host of nucleosynthesis models available in the literature. For some of the initial such studies applied to the LMC remnants, see [164]; and as a most recent example of a population systematic study of Galactic SNRs hosting CCOs and magnetars, while making use of a suite of models for core-collapse SNRs, see [44,413]. In many cases, the type of remnant remains unknown or controversial (e.g. [215,318,320,334,398,412]); this is in big part due to the lack of sensitive spatially resolved spectra needed to map the ejecta across the remnant (often on arcseconds scale) and to remove the degeneracy from the various nucleosynthesis model predictions. Fig. 25 and Fig. 26 illustrate two such examples of SNRs with a controversial type where *AXIS* will provide spectra of small-scale regions across remnants with a much better sensitivity than existing missions [358] which will enable us to a) reduce significantly the uncertainties on the abundances inferred to compare to the nucleosynthesis models, and b) remove models degeneracies associated with the comparison of fitted abundances to the different nucleosynthesis models.

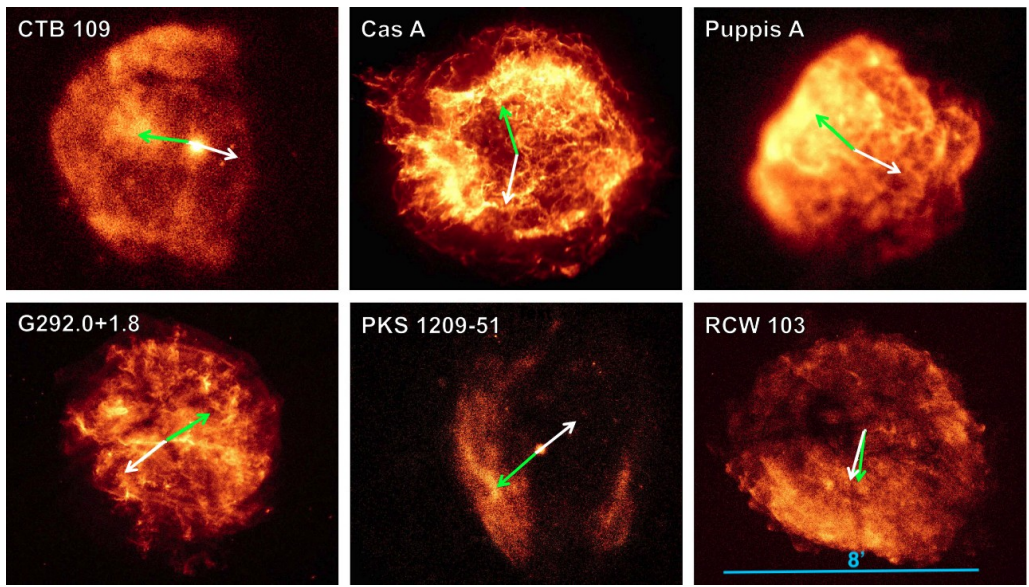

**Figure 27.** 0.5--2.1 keV *Chandra* and *ROSAT* images of six SNRs for which there is a robust measure of their explosion sites. The green arrow points from the explosion site to the direction of the dipole moment. The white arrow points in the direction of the neutron star's motion. In Cas A and G292.0+1.8, the dipole moment direction reflects the bulk of ejecta emission. In CTB 109, Puppis A, and RCW 103, the dipole moment points toward enhanced emission that is due to interactions with CSM/ ISM or a molecular cloud. The scale shown in the last panel shows the SNR size which can easily fit within the 24′-circular diameter FoV of AXIS. Figure adapted from [157]. See also Fig. 1 in [188].

In a systematic study of SNRs in the Magellanic Clouds with *XMM-Newton*, a new class of evolved SNRs were discovered, which have an X-ray bright, Fe-rich core, consistent with reverse shock-heated Type Ia ejecta [40,189,190]. As the interior of such SNRs show bright Fe L emission around 1 keV, the X-ray color can be used to determine the type of the SN not only in the nearby SNRs, for which one can perform studies of the morphology and the spectrum, but also in SNRs in further distant galaxies.

### 8.2. Probing Neutron Star kicks and their SN explosion mechanism

The origin of neutron stars' velocities (several hundreds km/s) is a long-standing mystery in astrophysics. There are two main competing mechanisms to neutron star kicks: (a) anisotropic ejection of the stellar debris (hydrodynamic kick, [171]) and (b) asymmetric-neutrino emission (neutrino-induced kick, [31,393]). Fortunately, the two scenarios predict a clear difference in the kick velocities and SN asymmetries. The hydrodynamic kick mechanism predicts that neutron star velocities are directed opposite to the stronger explosion where explosively nucleosynthesis elements from Si to Fe are preferentially expelled, whereas the neutrino-induced kick mechanism either suggests no correlation between neutron star velocities and SN asymmetries, or predicts the strongest mass ejection in the direction of the neutron star's motion.

Recent X-ray observations of SNRs revealed that neutron stars preferentially move opposite to the bulk of either X-ray emission [157] or intermediate-mass elements [188], supporting the hydrodynamic kick scenario. However, in many cases, neutron stars' velocities are indirectly inferred from displacements between their positions and geometric centers. The number of robust samples is still quite small and the sample includes a couple of outliers (see Fig. 27). Of the NSs associated with young Galactic SNRs with thermal emission from ejecta—necessary to distinguish between the two origin scenarios—only

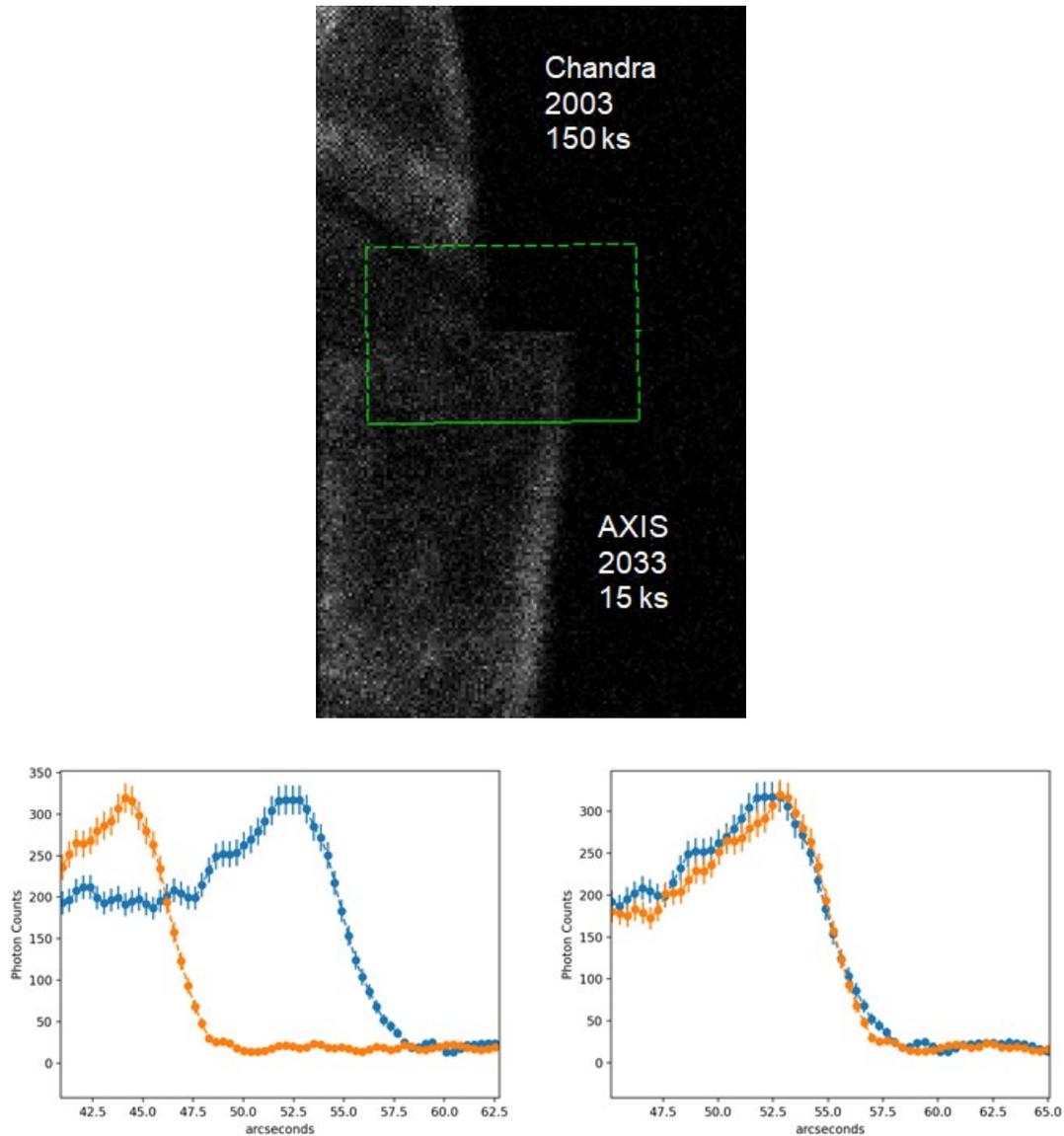

**Figure 28.** The edge of the Tycho SNR as seen in the 150 ks *Chandra* observation from 2003 (top) and a simulated 2033 15 ks *AXIS* observation (bottom part of the image). The radial profiles of the shell edge are shown in the bottom panel to illustrate how *AXIS* will be able to easily discern the proper motion in the 2030s. Orange represents the Chandra data and blue is the *AXIS* simulation. The left plot is the two profiles as measured, and the right plot is the same two curves but with the two curves shifted in order to align the profiles. In both plots the two profiles have been normalized to have the same profile height.

~7 of them have constraining direct proper motion measurements. *AXIS* will significantly increase the observational sample and is the most suitable mission suitable for tackling this science.

Firstly, *AXIS* - thanks to its unprecedented sensitivity and imaging resolution, will discover new CCOs (keeping in mind that CCOs were discovered thanks to *Chandra*'s imaging resolution). Secondly, *AXIS* will enable tracking the CCOs' proper motions over the *Chandra-AXIS* baseline for the *Chandra*-observed remnants. Thirdly, *AXIS* will do better astrometry than *Chandra* which is needed to constrain the neutron stars' proper motion. The larger FoV of *AXIS* will enable the robust detection of many background point sources that can be used to calibrate the astrometry of *AXIS* observations and obtain precise NS location measurements. Fourth, due to its larger FoV and low off-axis angular resolution, *AXIS* will enable the measurement of additional SNR explosion sites—and thus indirect NS velocities—through back evolving numerous ejecta filaments to a central position. Some of the most precise NS velocity measurements have come from this method (e.g., [135,355]), as this technique avoids the difficulties in measuring the small angular motions of NSs. Lastly, *AXIS* will prove long-baseline confirmations on the proper motions of NSs already measured. Multiple NSs have had their velocities revised to significantly lower values when using longer baselines (e.g., [228]), and determining the true fraction of SNR-embedded NSs with high (>1000 km s$^{-1}$) velocities is essential for informing simulations and understanding CCSNe explosion processes. The *Chandra-AXIS* (≥30-year) baseline will provide the timescale needed to constrain the proper motion of neutron stars, while taking advantage of their matched angular resolution. At the same time, the unprecedented sensitivity of *AXIS* combined with its high angular resolution over the full SNRs hosting the compact objects will enable a much better constraint on the bulk of X-ray emitting ejecta, including the ability to capture and measure bulk ejecta properties of larger SNRs than can be observed with *Chandra*.

### 8.3. Deciphering the SN environments through their shocks' proper motion

An often overlooked importance of the *Chandra* X-ray observatory is the ability to perform X-ray proper motion measurements at sub-arcsecond scales. The X-ray emission from fast moving shocks traces their forward edge. Tycho's supernova occurred in 1572 and its remnant has been observed regularly by *Chandra*. Characterizing the medium into which the blast wave is expanding provides insight into the conditions of the progenitor system prior to the explosion. This work has already begun by separating the existing *Chandra* data into different epochs and searching for changes in the proper motion, e.g., see [69] for the proper motion study of Kepler's supernova remnant. However, the degradation of *Chandra*'s low energy sensitivity makes the necessary exposures for new observations and studies of additional remnants prohibitively long. *AXIS* provides an opportunity to couple the existing high resolution *Chandra* data with new high quality observations yielding 30 plus year baselines. As a showcase of this method, we simulated a pair of 10 ks *AXIS* observations of the Tycho SNR based on the longest individual *Chandra* observation segment (176 ks). The two simulated images were then shifted by 0.654$''$ which represents the expected proper motion of the edge of the remnant in a two-year span. Treating the simulated images as real data, we extracted a radial profile of the edge of the remnant in each data set and calculated the required shift in order to re-align the data sets. The measured shift of 0.65$''$ is in agreement with the input value. To prove the ability to utilize both *Chandra* and *AXIS* observations, we simulated a 15 ks *AXIS* image based on the previous *Chandra* observation segment from 2009 and shifted the resulting image by 7.2$''$. See Fig. 28. The 150 ks *Chandra* observation segment from 2003 was then re-binned to match the predicted *AXIS* pixel size. The two resulting images correspond to a 30-year baseline between the *Chandra* observation and the simulated *AXIS* observation. The input shift corresponds to 9$''$ assuming constant expansion velocity. Our measured shift is 8.9$''$ ± 0.2$''$. These two simulations show that *AXIS* will provide year to year proper motion measurements of the fast shocks in young SNRs in a fraction of the time required by *Chandra*. In addition, new observations may be combined with existing data to produce valuable long baselines.

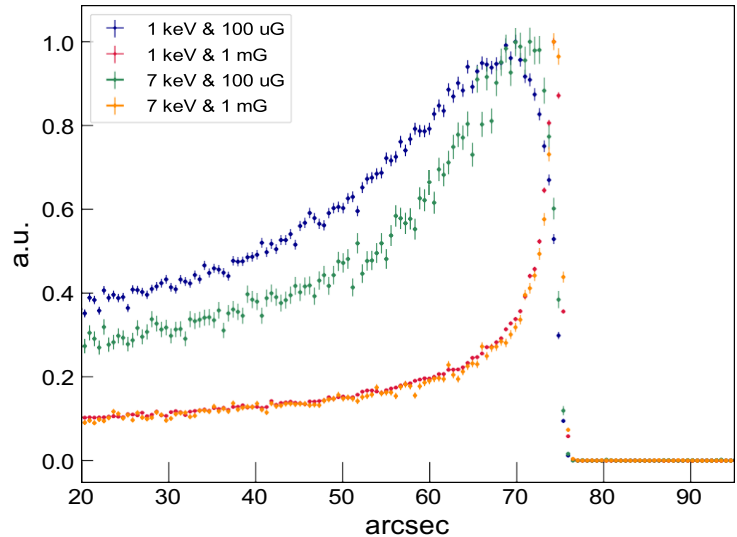

**Figure 29.** Simulated radial profiles of shocked emission from a young, 20 pc diameter SNR, at a distance of 1 kpc. The 100 ks *AXIS* simulations, for amplified magnetic fields of 100μG and 1 mG, illustrate that the shock narrows in width for stronger fields.

### 8.4. Probing magnetic field amplification in supernova remnant shocks

Turbulent magnetic fields at young SNR shocks are expected to be significantly amplified by a cosmic-ray current driven instability that develops in the shock precursor. Magnetic field amplification (MFA) is thought to be the key element in non-linear Diffusive Shock Acceleration (DSA) theory (see e.g., [47]). X-ray observations with *Chandra* have revealed the presence of narrow synchrotron X-ray filaments at the outer edge of young SNRs, demonstrating that strong shocks at young SNRs are indeed capable of amplifying the interstellar magnetic field by large factors (e.g., [375]). The narrowness of synchrotron X-ray filaments could be due to rapid synchrotron cooling of high-energy electrons in the postshock flow if the magnetic field strength reaches $\sim 0.1$ mG. In some cases, time variability of the non-thermal, synchrotron dominated X-ray hot spots can be seen (again using *Chandra*), which is additional evidence in favor of magnetic field amplification [361,362].

The turbulent magnetic fields, amplified most likely by cosmic ray current driven instabilities, can be imprinted in the spatial structures of the synchrotron X-ray filaments. Testing the theoretical predictions is best achieved with high-angular resolution measurements of the energy dependence on the width of the filamentary structures. The width of the shell $w$ depends on the magnetic field $B$, the energy of electron $E_e$ and the corresponding synchrotron X-ray photon energy $\varepsilon_X$: $w \propto E_e^{-1/2}B^{-7/4} \propto \varepsilon_X^{1/4} B^{-3/2}$ [417]. Figure 29 shows the radial profiles of the synchrotron X-ray filaments that are expected to be observed by *AXIS*. We can measure the energy dependence on the width of the synchrotron filaments thanks to the superb angular resolution of *AXIS* over a large field of view, thereby verifying the theoretical prediction. *AXIS* has the advantage of maintaining excellent angular resolution over the full SNR, while *Chandra* can generally only retain arcsecond spatial resolution over a small portion of the remnant.

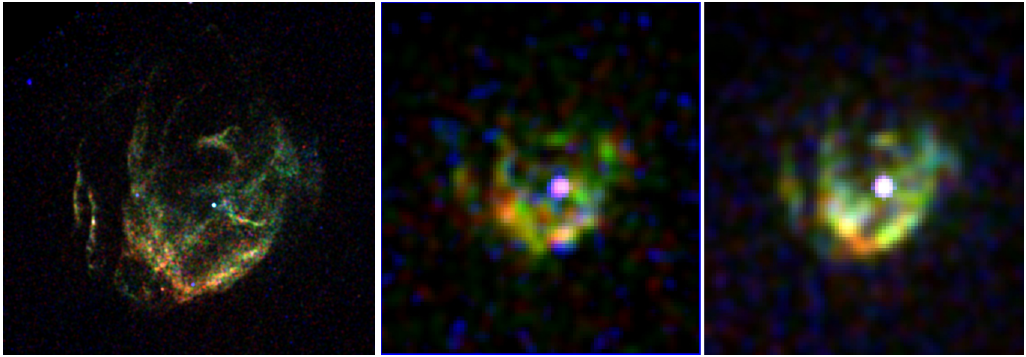

**Figure 30.** Left: *Chandra* observation of the Galactic SNR Kes 79 with a CCO (red: $0.5 - 1.2$ keV, green: $1.2 - 2.0$ keV, blue: $2.0 - 7.0$ keV). Middle: Image from a SIXTE simulation of an *AXIS* observation of Kes 79 assuming LMC distance and an exposure of 10 ks. Right: Same as in the middle image but for 50 ks.

### 8.5. Probing SNRs in the Magellanic Clouds

The largest and actively star-forming satellite galaxies of the Milky Way, the Large and Small Magellanic Clouds (LMC, SMC) are ideal places to study SNRs due to their low foreground absorption and small distances of ∼50 kpc and 62 kpc, respectively [119,365]. The proximity enables spatially resolved studies of the SNRs, and the accurately known distances permit the analysis of the energetics of each SNR. Observations of the entire SNR population in a galaxy on the other hand allow us to study the physics of SNRs, their progenitors, and the impact of the environment of SNRs on their evolution. A luminosity distribution of the SNRs in each galaxy can be constructed, which reveals the cause for the systematic offsets and slope differences between different galaxies, most likely caused by different overall abundances, average ISM densities, or typical explosion energies, in addition to selection effects. In addition, observations of the SNRs in the LMC and the SMC allow us to study not only the global properties of SNRs in a galaxy, but also specific sub-classes in detail (e.g., sorted by X-ray and radio morphology, diameter, or progenitor supernova type).

*Chandra*'s high-imaging resolution has provided us with an unprecedented view of the SNRs in the LMC and SMC. In particular, *Chandra* has played a pivotal role in (a) resolving and imaging young, metal-rich ejecta and their interaction with their surroundings [216,268,327,387], (b) zooming-in on the interiors of core-collapse SNRs while hunting for signs of their collapsed cores, PWNe or candidates [124,146, 213,221,376], and (c) measuring the expansion of their shock waves into the ISM [127,163,394]. Taking advantage of the high sensitivity of *XMM-Newton* especially in the softer band, a survey of SNRs and candidates in the Magellanic Clouds has been performed with *XMM-Newton* [e.g., 189,190,219,220]. In these studies, a new pulsar wind nebula or an X-ray binary in an SNR were discovered (e.g., LMC SNR J0453–6829 [133], LMC SNR J0513–6724 [222], SMC SNR DEM S5 [10]).

*AXIS* will allow us to continue the studies of the SNR populations in the Magellanic Clouds started with *Chandra* and *XMM-Newton*, with significantly improved sensitivity and spatial resolution over the full SNR. It will allow us to study each of the SNRs in great detail by resolving interior structures, and also detecting neutron stars or PWNe formed in their supernovae. As CCOs and AXPs are fainter X-ray sources with luminosities of $L_X = 10^{32-36}$ erg s$^{-1}$ (in comparison to XRBs, which are typically brighter $L_X \geq 10^{37}$ erg s$^{-1}$), only a handful of them have been detected in the Magellanic clouds SNRs (e.g., [219,220]). The combined resolution and unprecedented sensitivity of *AXIS* will allow us to resolve or detect a CCO with a modest exposure time even at the distance of the Magellanic Clouds (see Fig. 30).

## 9. Outflows from binaries

### 9.1. Microquasar jets

Microquasars are miniature versions of quasars, believed to host a central black hole surrounded by an accretion disk, and emitting relativistic jets that are very similar to those found in active galaxies. These jets offer an exciting probe of outflows from black holes and their feedback on the interstellar and intergalactic medium. Microquasar jets studies further provides a zoomed-in version of the large scale jets from quasars. The study of their dynamical evolution is possible on timescales that are inaccessible for active galactic nuclei jets.

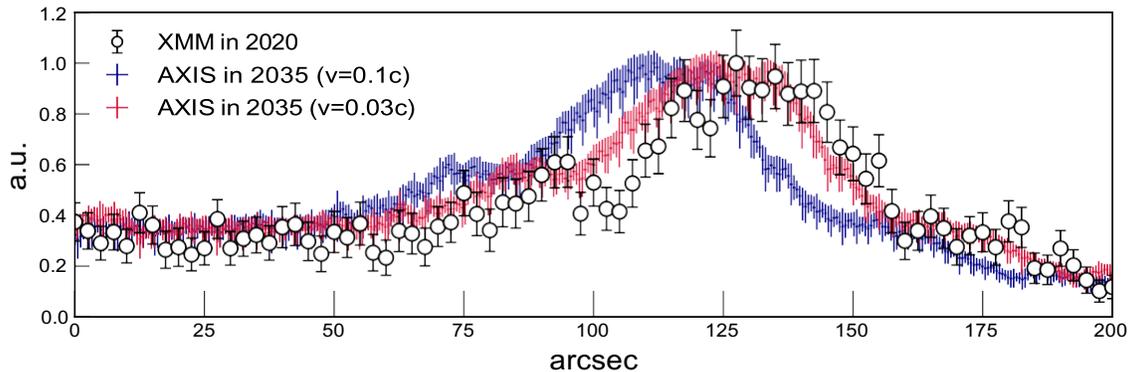

**Figure 31.** This figure illustrates the need for *AXIS* to follow the proper motion of microquasar jets (SS 433/W50 in this special case) while taking advantage of the legacy of *Chandra* and *XMM-Newton* observations of such outflows.

We so far know of radio/X-ray jets in five microquasars, and thanks to *Chandra*'s imaging resolution, these jets have been resolved into knots of emission that trace the ejection and particle acceleration process. The broadband spectrum of some of these is consistent with synchrotron emission from high-energy (up to 10 TeV) particles that were accelerated in the shock waves formed within the relativistic ejecta or by the interaction of the jets with the ISM. However often, the nature of the spectrum or scale size of the knots can not be confirmed given the poor count statistics or the decline of the PSF off-axis. *AXIS*'s unprecedented sensitivity and PSF over a large field of view will provide a leap in such studies and help discover new compact and large scale jets from a variety of accretion-powered sources.

One notable source known to power both short- and large-scale X-ray jets is the microquasar SS 433 powering the radio lobes of the large diffuse nebula, W50, recently detected in high-energy gamma-rays [7]. An X-ray observational campaign has led to the discovery of knots of hard X-ray emission that soften as we move away from SS 433 towards the large-scale lobes marking the site of the interaction between the jets and the ambient medium or SNR shell. The *Chandra* and *XMM-Newton* observations of this source shows that these knots, located at ≥30 pc away from SS 433, exhibit dynamic motion. However, the precise trajectories come with significant uncertainties. Notably, one of these knots serves as the location for particle re-acceleration to energies reaching up to a few hundred tera-electronvolts (TeV). The mechanism behind this re-acceleration remains poorly understood at present. In Fig 31, we show how *AXIS* will enable an accurate measurement of the velocity, under two velocity assumptions (0.03c and 0.1c). This will in turn address the physics of particle acceleration or deceleration as the large-scale jet encounters the surrounding medium (see [321] and references therein for a recent study).

*9.2. Transient Outflows*

When X-ray binaries undergo an outburst they can launch jets, typically observed at radio wavelengths. As these jet ejections fly away from the binary at mildly to highly relativistic speeds they eventually begin to interact with the ISM at late times, forming external shocks and accelerating particles, leading to a brightening in X-rays. These outflows are often large ($\sim 10''$) and relatively faint (see Figure 32), thus many can only be studied with *AXIS* due to its high sensitivity and exquisite angular resolution. Jet ejections from several BH X-ray binaries have been observed by *Chandra* [71,72,92,176,233]. Studying the late time deceleration and X-ray emission from these jets can help us to place tight constraints on their energetics, and therefore, possible launching mechanisms.

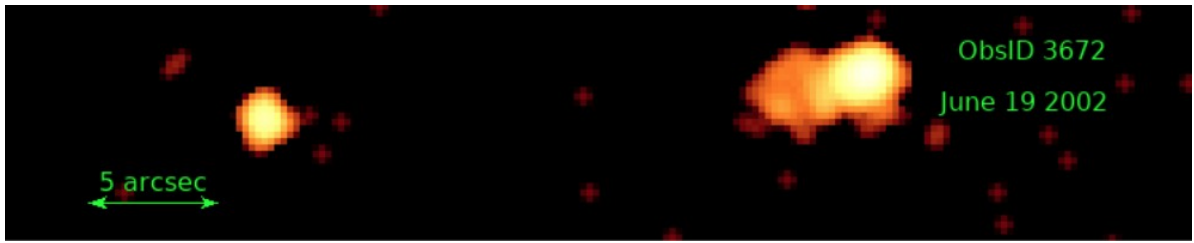

**Figure 32.** *Chandra* ACIS observations of the X-ray emitting jets launched by XTE J1550-564. The jet is the extended X-ray emission seen on the right side of the figure, while the binary is the X-ray source on the left. Figure adopted from [233] and modified.

Other X-ray binaries have also showed mildly relativistic outflows. For example, *Chandra* has observed X-ray emitting clumps being launched from the high mass gamma-ray binary LS 2883/PSR B1259-63, which move at velocities of $\approx 0.1c$ and show signs of being accelerated [141,143,183,270,271]. This material is thought to be debris from the massive companion's stellar disk, which is inclined to the orbit of the pulsar such that the pulsar passes through it twice each orbit. When the pulsar passes through the disk some of this material is ejected from the binary, and then is shocked and accelerated by its interaction with the pulsar wind (see [28,271] for a more detailed explanation). To date, this remains the only high-mass gamma-ray binary to show such phenomena, though it has been searched for in others (see e.g., [181,256]). *Chandra* has played a pivotal role in the discovery and follow-up of transient outflows from a wide array of different types of binaries. *AXIS* is the only observatory in the near future with sufficient sensitivity and angular resolution to allow for the continuation of these types of studies.

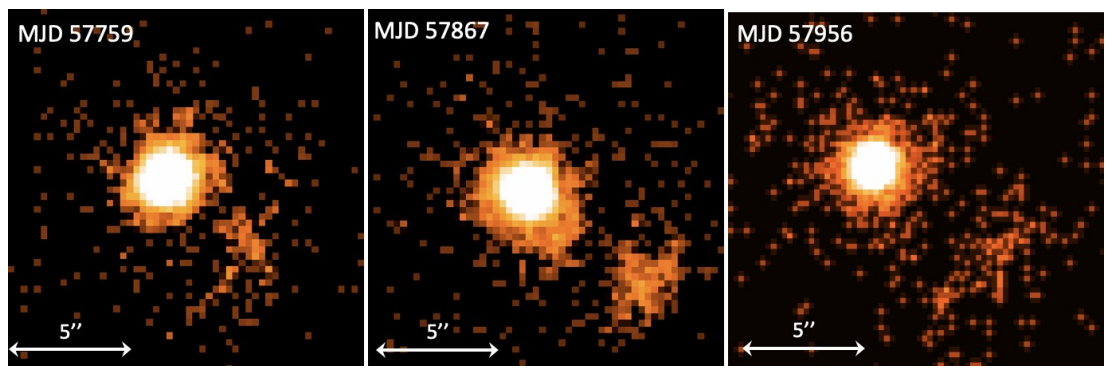

**Figure 33.** *Chandra* ACIS observations of the X-ray emitting clump launched from the high mass gamma-ray binary PSR B1259-63. The clump (south-west of the source) showed signs of acceleration and reached a velocity of $\approx 0.1c$ (see [141] for additional details).

## 10. Conclusions

Thanks to the *Chandra* X-ray Observatory, the past two decades showed us the power of high-resolution X-ray imaging enabling the discovery and study of new classes of compact objects, while zooming in on their relativistic outflows and their surrounding supernova remnants. We have only touched the tip of the iceberg, however, and more in-depth population studies and discoveries await the launch of *AXIS* thanks to its unprecedented sensitivity, relatively low background, and high-imaging resolution across its large field-of-view.

Following on the legacy of *Chandra*, and as we enter the era of high-resolution-imaging at other wavelengths, *AXIS* will enable a giant leap forward in X-ray astronomy, particularly in the study of compact objects and SNRs (both individual objects and population studies, in our Galaxy and in nearby galaxies). *AXIS* will:

- discover new types of (and some long sought) compact objects, magnetar and pulsar wind nebulae;
- enable population studies of the compact objects zoo and their connections to mysterious extragalactic sources (including Fast Radio Bursts and Gamma-ray Bursts);
- zoom-in on and track relativistic outflows and their interaction with the ISM;
- and peer into the physics of compact objects' kicks, supernova explosions, and their chemical and dynamical impact on the ISM.

In this white paper, we have surveyed the wealth of relevant topics to be pioneered by *AXIS* and to be made accessible to the X-ray community through the guest observer program, in synergy with multi-wavelength, multi-messenger observations for the next decade, including with *JWST*, Rubin, Roman, SKA pathfinders, CTA, and LISA.

**Acknowledgments:**

We express gratitude to all co-authors and endorsers who supported the earlier Astro2020 white paper [317], which underscored and endorsed the imperative, in the 2030s, for high-resolution X-ray imaging of neutron stars, pulsar wind nebulae, and supernova remnants. The realization of many scientific scenarios outlined in this white paper owes its feasibility to (a) the invaluable contributions from numerous members of the compact objects-PWN-SNR community (the list is extensive), (b) the capabilities and enduring impact of the Chandra X-ray Observatory [386], along with the dedicated efforts of its staff and the availability of user-friendly software tools, and (c) the inspiration and encouragement by the emerging generation of scientists and the public, urging the sustained exploration of the high-energy universe with sharp X-ray vision, especially within the era of multi-wavelength multi-messenger astrophysics.

S.S.H., A.M.M., L.M., C.T., C.B., and A.C. acknowledge support from the Natural Sciences and Engineering Research Council of Canada (NSERC) through the Canada Research Chairs program and the Discovery Grants program, the Canadian Institute for Theoretical Astrophysics (CITA), and the Canadian Space Agency (CSA). LM acknowledges partial salary support from a CITA National Fellowship (NSERC funding reference #DIS-2022-568580). H.A. acknowledges support from the National Research Foundation of Korea (NRF) grant funded by the Korean Government (MSIT) (NRF-2023R1A2C1002718). J.H. acknowledges support from NASA under award number 80GSFC21M0002. N.S. acknowledges the support from NASA (grant number 80NSSC22K0332), NASA FINESST (grant number 80NSSC22K1597), Columbia University Dean's fellowship, and a grant from the Simons Foundation. N.S. performed part of this work at the Aspen Center for Physics, which is supported by the National Science Foundation grant PHY2210452. G.V. acknowledges support by H.F.R.I. through the project ASTRAPE (Project ID 7802). J.T.L. acknowledges the support from NSFC through the grant E3GJ251110. O.K. acknowledges the support from NASA award 80NSSC21K0024 and Chandra Awards AR3-24003A and GO2-23048X issued by the Chandra X-ray Center operated by the Smithsonian Astrophysical Observatory.